\documentclass[american,superscriptaddress,twocolumn]{revtex4-1}
\usepackage[T1]{fontenc}
\usepackage[latin9]{inputenc}
\usepackage{color}
\usepackage{babel}
\usepackage{longtable}
\usepackage{float}
\usepackage{textcomp}
\usepackage{amsmath}
\usepackage{amssymb}
\usepackage{graphicx}
\usepackage{esint}
\usepackage[unicode=true,pdfusetitle,
 bookmarks=true,bookmarksnumbered=false,bookmarksopen=false,
 breaklinks=true,pdfborder={0 0 1},backref=false,colorlinks=true]
 {hyperref}

\makeatletter

\newcommand{\lyxmathsym}[1]{\ifmmode\begingroup\def\b@ld{bold}
  \text{\ifx\math@version\b@ld\bfseries\fi#1}\endgroup\else#1\fi}

\providecommand{\tabularnewline}{\\}

\usepackage{amsmath}
\let\oldAA\AA
\renewcommand{\AA}{\text{\normalfont\oldAA}}

\let\oldtextdegree\textdegree
\renewcommand{\textdegree}{\text{\oldtextdegree}}

\makeatother

\begin{document}
\title{
Temperature-induced negative differential conductivity}
\author{M. J. E. Casey}
\affiliation{College of Science and Engineering, James Cook University, Townsville
4810, Australia}
\author{D. G. Cocks}
\affiliation{College of Science and Engineering, James Cook University, Townsville
4810, Australia}
\affiliation{Present address: Research School of Physics and Engineering, The Australian
National University, Canberra 0200, Australia}
\author{G. J. Boyle}
\affiliation{College of Science and Engineering, James Cook University, Townsville
4810, Australia}
\affiliation{Present address: Linear Accelerator Technologies (FLA), Deutsches
Elektronen-Synchrotron (DESY), Notkestra{\ss}e 85, 22607 Hamburg,
Germany}
\author{M. J. Brunger}
\affiliation{College of Science and Engineering, Flinders University, GPO Box 2100,
Adelaide, SA 5001, Australia}
\author{S. Dujko}
\address{Institute of Physics, University of Belgrade, Pregrevica 118, 11080
Belgrade, Serbia}
\author{J. de Urquijo}
\affiliation{Instituto de Ciencias F{\'i}sicas, Universidad Nacional Aut{\'o}noma de M{\'e}xico,
62251, Cuernavaca, Morelos, M{\'e}xico}
\author{R. D. White }
\affiliation{College of Science and Engineering, James Cook University, Townsville
4810, Australia}
\email{Ronald.White@jcu.edu.au}

\begin{abstract}
We report on the existence of temperature induced negative differential
conductivity (NDC) for electrons in gaseous nitrogen. The important
role of superelastic rotational collisional processes in this phenomenon
is highlighted. A model cross-section set, utilised to ensure an accurate
treatment of superelastic processes and achieve thermal equilibrium
is detailed, and used to illustrate the role of de-excitation processes
in NDC. The criterion of Robson \citep{Robs84} for predicting the
occurance of NDC using only knowledge of the collisional cross-sections
is utilised for both the model system and N$_{2}.$ We also report
on the impact of anisotropy in the very low threshold scattering channels
on the transport coefficients, examine the Frost-Phelps finite difference
collision operator for the inelastic channel, in particular its neglect
of recoil, and assess other assumptions utilised in existing Boltzmann
equation solvers. We discuss the numerical challenges associated with
low reduced electric field calculations, and detail an alternative
representation of the elastic and inelastic collision operators used
in Boltzmann equation solutions that enforce conservation of number
density. Finally, new experimental measurements of the drift velocity
and the Townsend ionisation coefficient for an electron swarm in N$_{2}$
are reported from a pulsed Townsend experiment. The self-consistency
of the utilised cross-sections is also briefly assessed against these
results.
\end{abstract}
\maketitle

\section{Introduction\label{sec:Introduction}}

As a swarm of electrons drift and diffuse through a background medium,
driven out of equilibrium by an externally applied electric field,
there can exist a region where the drift velocity of the electrons
decreases with increasing electric field strength. This phenomenon,
known as negative differential conductivity (NDC), has been comprehensively
studied, both experimentally and theoretically~\citep{Petrovic1984,Dyatko2014}.
In both plasma and swarm physics NDC is present in gases used for
dosimetry and particle detectors~\citep{Christophorou1979,Mathieson1979,AlDargazelli1981},
and has implications on the operating ranges of gas lasers with NDC-induced
electric current oscillations in electron-beam-sustained discharge
switches~\citep{Lopantseva1979,Christophorou1981}. For fundamental
physics, NDC has played a role in evaluating complete and accurate
scattering cross-section sets~\citep{Petrovic2007,Nakamura1995}.
Argon, for example, was considered to be a candidate for NDC in a
pure gas, but this was later shown to be due to the presence of molecular
impurities in Ar samples~\citep{Long1976}, and in gaseous mercury
it has recently been shown that NDC occurs due to the presence of
dimers~\citep{Miric2017}. Swarms of electrons can also induce NDC
in liquids~\citep{WhitRobs09}, plasmas and semiconductors~\citep{Chiflikian1995,Chiflikyan2000,Dyatko2007},
and NDC has been shown to be induced by positron swarms in argon~\citep{Suvakov2008}.
In addition to NDC for positrons in argon, the same phenomenon has
been observed in water vapour~\citep{Bankovic2012}, molecular hydrogen~\citep{Bankovic2012b}
and CF$_{4}$~\citep{Bankovic2014}. As such, modelling systems to
predict regions of NDC and the conditions of electron-induced NDC
is of particular interest~\citep{Robs84,Petrovic1984}.

In model systems, the work of Petrovi{\'c} \emph{et al.}~\citep{Petrovic1984}
and Vrhovac and Petrovi{\'c}~\citep{VrhoPetr1996} detail different
systems involving elastic, inelastic and ionisation cross-sections
that, under various conditions, either enhance or eliminate the occurrence
of an NDC region. Further to the electron and position experimental
studies, in real systems, the existence of NDC has been observed experimentally
in gaseous systems of N$_{2}$~\citep{PackPhelps_lx}, CH$_{4}$,
CF$_{4}$~\citep{Hunter1985} and Hg~\citep{Miric2017}, and predicted
theoretically due to electron-electron interactions (\citep{Chiflikyan2000})
in plasmas of Xe~\citep{Aleksandrov1996,DonkoDyatko2016}, to name
but a few. NDC in gas mixtures also has an extensive history, observed
in mixtures with helium, argon, N$_{2}$ and CH$_{4}$~\citep{Chiflikian1995,Hunter1985,Dyatko2010,Dyatko2014,Stano2011}.
In strongly attaching gases, NDC has also been shown to be induced
through a combination of attachment heating and inelastic cooling~\citep{Miric2016}.

Throughout this broad body of work, the sources of NDC have been discussed
in detail by the various authors and a number of criteria have been
proposed. Some of the conditions under which NDC can occur include
the presence of inelastic collision channels, favoured particularly
by a decreasing inelastic cross-section, the presence of a Ramsauer-Townsend
minimum in the elastic momentum-transfer cross-section, or a rapidly
increasing elastic momentum-transfer cross-section. However these
are not necessary and sufficient conditions. The validity of these
early criteria on the understanding of NDC was discussed in detail
in Petrovi{\'c} \emph{et al.}~\citep{Petrovic1984} who address
the analyses of Kleban and Davis~\citep{klebanDavis1977,klebanDavis1978},
Long and co-workers~\citep{Long1976}, and Lopantseva and co-workers~\citep{Lopantseva1979},
in particular, and in Vrhovac and Petrovi{\'c}~\citep{VrhoPetr1996}
where consistency with the Shizgal~\citep{Shizgal1990} criterion
is discussed. Of particular note are the simulations of Petrovi{\'c}
\emph{et al.}~\citep{Petrovic1984} for N$_{2}$ that confirm that
the presence of inelastic processes, other than rotational excitations,
decreases the range of the NDC region, indicating that the rotational
collisions are responsible for the presence of NDC, where the elastic
cross-section is relatively isotropic. This is explored further in
this work, where the effect of temperature on the presence or absence
of NDC is explored. The thermally induced NDC region in N$_{2}$ below
room temperature is detailed. Here, the contribution of inelastic
ground-state and excited-state collisional processes to the nett energy
transfer to and from the electron swarm were shown to be responsible
for the extent of an NDC region.

The various criteria for the presence of NDC, detailed in these early
works, as noted above, have been discussed in detail by the respective
authors. Of particular interest here is the criterion proposed by
Robson~\citep{Robs84}, where momentum-transfer theory was used to
derive an expression based on the rate of change of the ratio of inelastic
to elastic energy transfer with energy. This criterion allows prediction
of NDC using only a knowledge of the collisional cross-sections, the
accuracy of which is highlighted using both a simple model system
and for N$_{2}$. The thermally-induced reduction and deactivation
of NDC is explored further using Robson's criterion.

A model collisional system is used throughout this study to simplify
discussions around NDC and its temperature dependence. The model system
is also used to verify both our solution method, and explicitly test
the inclusion of temperature dependent inelastic collisions. This
system also facilitates further discussion around some of the assumptions
sometimes involved in swarm modelling. The explicit effect of anisotropy
in low-threshold inelastic processes is assessed, as well as the commonly
employed two-term approximation~\citep{White2003a}, as is the neglect
of superelastic collisions in higher-order collisional terms. Also
assessed is the neglect of the recoil of the neutral particle during
inelastic collisions in the excitation collision operator, as utilised
in many Boltzmann equation solutions. Truncation of the mass ratio
expansion at zeroth order for inelastic collisions is compared with
the exact collision description of Monte Carlo simulations, for very
low-energy threshold processes like rotational excitations, where
the energy exchange is much closer to that for elastic collisions.
For application of this discussion to real gases, these assumptions
are also assessed for electron swarms in N$_{2}$. For the low energy
regime of interest in this study the discrete rotational collisions
in N$_{2}$ are treated using the Frost-Phelps collision operator,
although in the work of Ridenti \emph{et al.}~\citep{Ridenti2015}
the Chapman-Cowling extension to the continuous approximation to rotations
was developed to bridge the continuous energy loss regime applicable
at high fields to the discrete collision description for use at low
energies.

The particular numerical techniques and challenges associated with
modelling low-threshold processes, in low electric field conditions,
are discussed for both our model system and in N$_{2}$. For a robust
solution method to reproduce various experimental conditions, standard
collisional benchmarks are employed along with self-consistency checks
of numerical number density conservation, energy and momentum balance.
Under particular conditions we are able to provoke non-physical solutions.
The effect of these problematic solutions is assessed and addressed
using alternative representations of the standard collision operators,
and compared with the results of an independent Monte Carlo code. 

This work is arranged by first detailing our kinetic theory, our multi-term
solution and particular numerical approaches, in section~\ref{sec: Theory}
and Appendix~\ref{sec: Appendix Numerical}. The Monte Carlo code
used for comparison with our results is briefly outlined in section~\ref{subsec: Monte-Carlo-Technique},
while the pulsed Townsend apparatus, used for measurement of the drift
velocity and the Townsend ionisation coefficient in N$_{2}$, is also
briefly detailed in section~\ref{subsec: Experimental-technique}.
In section~\ref{sec:Results-and-Discussion} we consider two systems
that exhibit NDC---a simple model system in section~\ref{subsec:Benchmark-model}
and N$_{2}$ in section~\ref{subsec:N2}. For both systems we present
calculations of transport coefficients at various temperatures, and
discuss the thermal activation of NDC and the necessity of de-excitation,
or superelastic, collisions for its occurrence. The prediction of
NDC using Robson's~\citep{Robs84} criterion from knowledge of the
energy transfer rates is presented for both systems, in sections~\ref{subsec: Criteria-for-NDC}
and \ref{subsec: NDC-in-N}. We also consider various physical and
numerical elements including: (i) the effect of the neglect of recoil
in the Frost-Phelps differential finite difference inelastic collision
operator, (ii) the two-term approximation on the  Townsend ionisation
coefficient, (iii) the effect of anisotropy in low-threshold inelastic
collision channels of our model cross-section, and (iv) the neglect
of higher order de-excitations sometimes applied in the solution of
similar problems, in sections~\ref{subsec: Model_approximations}
and \ref{subsec: N2_ApproximationEffects}. In section~\ref{subsec: Temperature-dependence-N2}
and Appendix~\ref{sec: ExpTransCo}, we also present new experimental
measurements of the drift velocity and the Townsend ionisation coefficient
for electron swarms in N$_{2}$ to compare against our calculations. 

\section{Theory\label{sec: Theory}}

\subsection{The Boltzmann equation and a multi-term solution framework}

The transport of a swarm of charged particles through a gaseous medium
is described by the particle phase space distribution function $f(\mathbf{r},\mathbf{v},t)$,
representing the distribution of electrons with position $\mathbf{r}$,
velocity $\mathbf{v}$ and time $t$, that is the solution of the
linear Boltzmann equation, 
\begin{equation}
\left(\frac{\text{\ensuremath{\partial}}}{\text{\ensuremath{\partial}}t}+\mathbf{v}\cdot\text{\ensuremath{\nabla}}+\frac{q\mathbf{E}}{m_{e}}\cdot\frac{\text{\ensuremath{\partial}}}{\text{\ensuremath{\partial}\ensuremath{\mathbf{v}}}}\right)f\left(\mathbf{r},\mathbf{v},t\right)=\text{\textminus}J(f\left(\mathbf{r},\mathbf{v},t\right)),\label{eq:Boltz}
\end{equation}
where $m_{e}$ is the mass of the swarm particle, $q$ is the elementary
charge, and $\mathbf{E}$ is the externally applied electric field.
The linear collision operator $J$ describes binary collisions between
the swarm particles and the background medium, and accounts for elastic
and inelastic collisions, and particle non-conserving loss (attachment)
and gain (electron impact ionisation) collisions.

For a solution to the Boltzmann equation, the angular dependence of
the phase space distribution function is expanded in terms of spherical
harmonics, to give:
\begin{equation}
f(\mathbf{r},\mathbf{v},t)=\sum_{l=0}^{\infty}\sum_{m=-l}^{l}f_{m}^{(l)}(\mathbf{r},v,t)Y_{m}^{[l]}(\mathbf{\hat{v}}).\label{eq: SphHarmExpans}
\end{equation}
In practice, the index $l$ must be truncated at some upper value
$l_{\textrm{max}}$, incremented until some convergence criterion
on the distribution function, or its velocity moments, is met. For
this work we do not restrict the truncation at $l_{\textrm{max}}=1$,
as is commonly done for the `two-term approximation', sometimes leading
to an inadequate representation of the anisotropic parts of the distribution
function and incorrect transport coefficients, see the review~\citep{White2003a}.
Substitution of the expansion into equation~\eqref{eq:Boltz} leads
to a system of coupled equations for $f_{m}^{(l)}$~\citep{RobsWinkSige02}. 

For the plane parallel geometry representing our experimental conditions,
the preferred direction is taken to be perpendicular to the electrodes
and the spatial gradients are along the $z$ axis so that $\mathbf{r}=z$
and $\mathbf{E}=E_{z}$, and the $m$ index is restricted to $m=0$
by symmetry. Equation~\eqref{eq:Boltz}, with substitution of the
expansion~\eqref{eq: SphHarmExpans} and recast in energy-space using
$qU=\frac{1}{2}m_{e}v^{2}$ for $U$ in eV, becomes:
\begin{align}
\begin{aligned} & \frac{\partial f_{l}}{\partial t}+J_{l}(f_{l})+\left(\frac{2q}{m_{e}}\right)^{1/2}\sum_{p=\pm1}\Delta_{l}^{(p)}\Bigg[U^{1/2}\frac{\partial}{\partial z}\\
 & +E_{z}\Bigg(U^{1/2}\frac{\partial}{\partial U}+\frac{p}{2}\left(l+\frac{3p+1}{2}\right)U^{-1/2}\Bigg)\Bigg]\,f_{l+p}=0,
\end{aligned}
\nonumber \\
\Delta_{l}^{(+)}=\frac{l+1}{2l+3},\nonumber \\
\Delta_{l}^{(-)}=\frac{l}{2l-1},\label{eq: BoltzPlaneParallel}
\end{align}
where $J_{l}$ is the Legendre decomposition of the collision operator,
detailed in the next section, and $E_{z}$ is the electric field defined
parallel to the $z$ axis. 

\subsection{Collision operators\label{subsec: Inel Coll Op}}

\subsubsection{Elastic collisions}

For electron swarms in atomic and molecular gases the small mass ratio
is utilised so that the Davydov operator for elastic collisions may
be used~\citep{Davy35,WhitRobs11,Pidduck1936}, and is given by:
\begin{eqnarray*}
J_{l}^{\textrm{elas}} & = & \begin{cases}
\begin{aligned} & -\frac{2m_{e}}{m_{0}}U^{-1/2}\frac{\partial}{\partial U}\Bigg[U^{3/2}\nu_{m}^{\textrm{elas}}(U)\\
 & \times\left(f_{0}(U)+\frac{k_{B}T_{0}}{q}\frac{\partial}{\partial U}f_{0}(U)\right)\Bigg]
\end{aligned}
 & l=0\\
\begin{aligned} & \nu_{l}^{\textrm{elas}}(U)f_{l}(U),\end{aligned}
 & l\geq1
\end{cases}
\end{eqnarray*}
where $\nu_{m}$ is the momentum-transfer collision frequency, $\nu_{l}$
is the $l$th partial collision frequency, in $\textrm{s}^{-1}$,
$k_{B}$ is Boltzmann's constant, and $T_{0}$ is the temperature
of the neutral background gas.

\subsubsection{Inelastic collisions}

The Frost and Phelps Legendre-decomposed collision operator~\citep{Frost1962}
is employed here to describe the effect of inelastic particle-conserving
collisions on the spatially-independent velocity distribution function.
The anisotropic form of the collision operator was detailed by Makabe
and White~\citep{MakabeWhite2015}, Phelps and Pitchford~\citep{PhelpsPitch1985_aniso}
and earlier in Reid~\citep{Reid1979} (in the second and third terms
on the right hand side of the equation following equation (3)), and
is here extended to include de-excitation, or superelastic, collisions
using detailed balance~\citep{Hochstim}. Expressed in terms of initial
and final internal states $j$ and $k$ of the neutral particle, where
$j<k$, particles with energy above the threshold $U_{th}$ are available
for excitations from $j\rightarrow k$. Below the threshold, for non-zero
temperatures, the background neutral particles may be in an excited
state $k$ and are available to undergo superelastic collisions from
$k\rightarrow j$, where the energy loss, taken to be the threshold,
is gained by the incoming electron in a de-excitation process and
lost by the neutral particle.

The partial cross-sections $\sigma_{l}$ are the coefficients of a
Legendre polynomial ($P_{l}$) expansion of the differential cross-sections,
$\sigma(U,\chi)$ for the scattering angle $\chi$, defined by $\sigma_{l}(U)=2\pi\int_{-1}^{1}\sigma(U,\chi)P_{l}(\cos\chi)d(\cos\chi)$.
The de-excitation cross-section $\sigma_{l}(kj;U)$ is expressed in
terms of the excitation cross-section $\sigma_{l}(jk;U)$ using the
microscopic reversibility relation $g_{k}U\sigma_{l}(kj;U)=g_{j}(U+U_{th})\sigma_{l}(jk;U+U_{th})$
where $g_{k}$ and $g_{j}$ are the degeneracy of the $k$th and $j$th
states. After converting to collision frequencies through $\nu_{l}(U)=n_{0}v\sigma_{l}(U)=n_{0}\sqrt{\left(2qU/m_{e}\right)}\sigma_{l}$
in energy space, where $n_{0}$ is the neutral number density, the
isotropic and anisotropic components of the inelastic collision operator
are given by:
\begin{eqnarray}
J_{0}^{\textrm{inel}} & = & \sum_{j,k}n_{0j}\Bigg\{\left(\frac{U+U_{th}}{U}\right)^{1/2}f_{0}(U+U_{th})\label{eq: InelasCollOp}\\
 &  & \times\nu_{0}(jk;U+U_{th})-f_{0}(U)\nu_{0}(jk;U)\Bigg\}\nonumber \\
 &  & +\sum_{j,k}n_{0k}\frac{g_{j}}{g_{k}}\Bigg\{ f_{0}(U-U_{th})\nu_{0}(jk;U)\nonumber \\
 &  & -\left(\frac{U+U_{th}}{U}\right)^{1/2}f_{0}(U)\nu_{0}(jk;U+U_{th})\Bigg\},\nonumber 
\end{eqnarray}
\begin{eqnarray}
J_{l\geq1}^{\textrm{inel}} & = & \sum_{j,k}n_{0j}\Bigg\{\left(\frac{U+U_{th}}{U}\right)^{3/2}f_{l}(U+U_{th})\label{eq: InelasCollOp2}\\
 &  & \times\nu_{l}(jk;U+U_{th})-f_{l}(U)\nu_{0}(jk;U)\Bigg\}\nonumber \\
 &  & +\sum_{j,k}n_{0k}\frac{g_{j}}{g_{k}}\Bigg\{\left(\frac{U-U_{th}}{U}\right)f_{l}(U-U_{th})\nonumber \\
 &  & \times\nu_{l}(jk;U)\nonumber \\
 &  & -\left(\frac{U+U_{th}}{U}\right)^{1/2}f_{l}(U)\nu_{0}(jk;U+U_{th})\Bigg\},\nonumber 
\end{eqnarray}
where $n_{0j}$ and $n_{0k}$ are the density of neutral particles
in the initial and final states $j$ and $k$, respectively, $\nu_{l}$
is the $l$th partial collision frequency, related to the momentum-transfer
cross-section through $\sigma_{m}=\sigma_{0}-\sqrt{\frac{U+U_{th}}{U}}\sigma_{1}$
for inelastic collisions. The number density of the neutral particles
in the state $j$ with energy $U_{j}$, are calculated using standard
Boltzmann statistics: $n_{0j}=\frac{n_{0}}{Z}g_{j}\exp\left(\frac{-U_{j}}{k_{B}T_{0}}\right)$
where the partition function sums over all possible internal states
$j$ and is given by $Z=\sum_{j}g_{j}\exp\left(\frac{-U_{j}}{k_{B}T_{0}}\right)$.

\subsubsection{Ionising collisions}

The electron-impact ionisation operator $J_{l}^{\textrm{ion}}$ utilised
here, in Legendre-decomposed form, is given by:
\[
J_{l}^{\textrm{ion}}(f_{l})=\begin{cases}
\begin{aligned} & \nu^{\textrm{ion}}(U)f_{0}(U)-2U^{-1/2}\\
 & \times\!\!\int_{0}^{\infty}\!\!\!dU'U'^{1/2}\nu^{\textrm{ion}}(U')P(U,U')f_{0}(U'),
\end{aligned}
 & \!\!l=0\\
\begin{aligned} & \nu^{\textrm{ion}}(U)f_{l}(U),\end{aligned}
 & \!\!l\geq1,
\end{cases}
\]
where $\nu^{\textrm{ion}}$ is the ionisation collision frequency,
and $P(U,U')$ is the energy-partitioning function~\citep{NessRob1986}.

\subsection{Solution technique\label{subsec: Solution-technique}}

\paragraph{Time-of-flight:}

When the number density varies slowly in space, away from boundaries
and under the influence of a uniform electric field, hydrodynamic
conditions prevail and the space-time dependence of the distribution
function can be projected onto the number density so that $f_{l}(z,U,t)=\sum_{s}f_{l}^{s}(U)\frac{\partial^{s}n(z,t)}{\partial z^{s}}$,
where $s$ is the rank of the tensor. 

To account for particle non-conserving processes, the density gradient
expansion to second order is required. In plane-parallel geometry
this is given by: 
\begin{eqnarray*}
f_{l}(z,U,t) & = & F_{l}(U)n(z,t)-F_{l}^{(L)}(U)\frac{\partial n(z,t)}{\partial z}\\
 &  & +\sqrt{\frac{1}{3}}F_{l}^{(2T)}(U)\frac{\partial^{2}n(z,t)}{\partial z^{2}}\\
 &  & -\sqrt{\frac{2}{3}}F_{l}^{(2L)}(U)\frac{\partial^{2}n(z,t)}{\partial z^{2}},
\end{eqnarray*}
where the $L$ and $T$ superscripts on the distribution $F_{l}$
are defined parallel and transverse to the electric field, respectively.
For weak gradients, a density gradient expansion of the phase-space
distribution function may be taken and the resulting diffusion equation~\citep{RobWhiteNess2011}:
\[
\frac{\partial n}{\partial t}+W\frac{\partial n}{\partial z}-D_{L}\frac{\partial^{2}n}{\partial z^{2}}=nR
\]
is used to analyse experimental parameters. The time-of-flight coefficients
of the density gradient expansion are found from the solution to the
hierarchy,
\begin{align}
 & \frac{\partial}{\partial t}\phi_{l}+\omega_{0}\phi_{l}+E_{z}\left(\frac{2q}{m_{e}}\right)^{1/2}\sum_{p=\pm1}\Delta_{l}^{(p)}\label{eq: BoltzHydro}\\
 & \times\left(U^{1/2}\frac{\partial}{\partial U}+\frac{p}{2U^{1/2}}\left(l+\frac{(3p+1)}{2}\right)\right)\phi_{l}+J_{l}\phi_{l}=h_{l}^{(s)}\nonumber 
\end{align}
where $\phi_{l}=\left\{ F_{l},\,F_{l}^{(L)},\,F_{l}^{(T)},\,F_{l}^{(2T)},\,F_{l}^{(2L)}\right\} $
and the $h_{l}^{(s)}$ are given by,
\begin{eqnarray*}
h_{l} & = & 0,\\
h_{l}^{(L)} & = & \left(\frac{2qU}{m_{e}}\right)^{1/2}\left(\frac{l+1}{2l+3}F_{l+1}+\frac{l}{2l-1}F_{l-1}\right)-\omega_{1}F_{l},\\
h_{l}^{(2T)} & = & \left(\frac{2qU}{3m_{e}}\right)^{1/2}\bigg[\frac{l+1}{2l+3}\left(F_{l+1}^{(L)}+(l+2)F_{l+1}^{(T)}\right)\\
 &  & +\frac{l}{2l-1}\left(F_{l-1}^{(L)}-(l-1)F_{l-1}^{(T)}\right)\bigg]\\
 &  & -\omega_{2}F_{l}-\left(\frac{1}{3}\right)^{1/2}\omega_{1}F_{l}^{(L)},\\
h_{l}^{(2L)} & = & -\left(\frac{qU}{3m_{e}}\right)^{1/2}\bigg[\frac{l+1}{2l+3}\left(2F_{l+1}^{(L)}-(l+2)F_{l+1}^{(T)}\right)\\
 &  & +\frac{l}{2l-1}\left(2F_{l-1}^{(L)}+(l-1)F_{l-1}^{(T)}\right)\bigg]\\
 &  & -\bar{\omega}_{2}F_{l}+\left(\frac{2}{3}\right)^{1/2}\omega_{1}F_{l}^{(L)}.
\end{eqnarray*}
The equation for the first level transverse distribution function
$F_{l}^{(T)}$ takes a different form: 
\begin{align*}
 & \left(J_{l}+\omega_{0}+\frac{\partial}{\partial t}\right)F_{l}^{(T)}\\
 & +\left(\frac{2q}{m_{e}}\right)^{1/2}\frac{l+2}{2l+3}E_{z}\left(U^{1/2}\frac{\partial}{\partial U}+\frac{l+2}{2}U^{-1/2}\right)F_{l+1}^{(T)}\\
 & +\left(\frac{2q}{m_{e}}\right)^{1/2}\frac{l-1}{2l-1}E_{z}\left(U^{1/2}\frac{\partial}{\partial U}-\frac{l-1}{2}U^{-1/2}\right)F_{l-1}^{(T)}\\
 & =\left(\frac{2qU}{m_{e}}\right)^{1/2}\left(\frac{1}{2l-1}F_{l-1}-\frac{1}{2l+3}F_{l+1}\right).
\end{align*}
Each of the expansion coefficients $F_{l}^{(s)}$ satisfy the normalisation
condition $2\pi\left(\frac{2q}{m_{e}}\right)^{3/2}\int_{0}^{\infty}U^{\frac{1}{2}}F_{0}^{(s)}dU=\delta_{s,0}$.
The coefficients $\omega$, using the density gradient expansion,
are given by:
\begin{eqnarray}
\omega_{0} & = & -2\pi\left(\frac{2q}{m_{e}}\right)^{3/2}\int U^{1/2}J_{0}^{R}\left(F_{0}\right)dU,\label{eq: omega_coeffs}\\
\omega_{1} & = & \frac{2\pi}{3}\left(\frac{2q}{m_{e}}\right)^{2}\int UF_{1}dU\nonumber \\
 &  & -2\pi\left(\frac{2q}{m_{e}}\right)^{3/2}\int U^{1/2}J_{0}^{R}\left(F_{0}^{(L)}\right)dU,\nonumber \\
\omega_{2} & = & \frac{2\pi}{3}\left(\frac{2q}{m_{e}}\right)^{2}\int U\left(F_{1}^{(L)}+2F_{1}^{(T)}\right)dU\nonumber \\
 &  & -2\pi\left(\frac{2q}{m_{e}}\right)^{3/2}\int U^{1/2}J_{0}^{R}\left(F_{0}^{(2T)}\right)dU,\nonumber \\
\bar{\omega}_{2} & = & -\frac{2\pi}{3}\left(\frac{2q}{m_{e}}\right)^{2}\int U\left(F_{1}^{(L)}-F_{1}^{(T)}\right)dU\nonumber \\
 &  & -2\pi\left(\frac{2q}{m_{e}}\right)^{3/2}\int U^{1/2}J_{0}^{R}\left(F_{0}^{(2L)}\right)dU,\nonumber 
\end{eqnarray}
where $J_{l}^{R}$ is the particle non-conserving, or reactive, collision
operator. In the presence of non-conservative collisions these coefficients
involve an integration over the density gradient expansion coefficients
$F_{l}^{(s)}$, so that these expressions become non-linear.

\paragraph{Steady-state Townsend:}

In modelling the steady-state Townsend experiment~\citep{Huxl40,Town15,HuxlCrom74},
for the steady-state time-asymptotic solution, far from the source
where no memory of the initial source distribution remains, the spatially
varying distribution function takes the form of a sum of exponentials
$f_{l}(z,U,t)=\psi_{l}(U)\exp(\omega t+kz)$, where $\omega$ and
$k$ are separation constants~\citep{RobWhiteNess2011}. Substitution
into equation~\eqref{eq: BoltzPlaneParallel}, leads to the generalised
eigenvalue equation:
\begin{align}
 & \omega\psi_{l}+\left(\frac{2q}{m_{e}}\right)^{1/2}\sum_{p=\pm1}\Delta_{l}^{p}\Bigg[U^{1/2}k\nonumber \\
 & +E_{z}\Bigg(U^{1/2}\frac{\text{\ensuremath{\partial}}}{\text{\ensuremath{\partial}}U}+\frac{p}{2U^{1/2}}\left(l+\frac{(3p+1)}{2}\right)\Bigg)\Bigg]\psi_{l+p}\nonumber \\
 & =-J_{l}\left(\psi_{l}\right),\label{eq: BoltzEvalue}
\end{align}
where the coefficients $\Delta_{l}^{p}$ are defined in equation~\eqref{eq: BoltzPlaneParallel}
and the eigenvalues $\omega$ and $k$ are related through the dispersion
relation $\Omega_{n}(\omega,k)=0,\;n=0,1,2,\dots$. 

For the steady-state Townsend experiment, the spatial eigenvalues
$k$ are found by setting $\omega$ to zero, which implies we are
looking at the temporally asymptotic regime where $\frac{\partial\psi_{l}}{\partial t}=0$.
The spatial eigenvalues $k$ are assumed to form a discrete set $k_{n}$
where the lowest magnitude non-zero eigenvalue $k_{1}$ represents
the reduced macroscopic Townsend ionisation coefficient $\alpha_{T}/n_{0}$.

\paragraph{Utility of the generalised eigenvalue method:\label{par: Generalised_Eigenvalue}}

The generalised eigenvalue equation~\eqref{eq: BoltzEvalue} may
be solved directly as an eigenvalue problem, the details of which
are omitted here but are described in Boyle~\citep{Boyle2015thesis}.
With this technique, the lowest remaining temporal eigenvalue of a
time-of-flight simulation represents the rate of non-conservative
processes in the long-time limit, equivalent to $\omega_{0}$ in equation~\eqref{eq: omega_coeffs},
and the corresponding eigenfunction represents the electron energy
distribution function. This provides an alternative method for calculating
the nett rate coefficient of non-conservative collisional processes
for a time-of-flight simulation that is typically calculated as the
integral of the non-conservative collision frequencies with the distribution
function, defined below in equations~\eqref{eq: TransCo}--\eqref{eq: TransCo1}.

\paragraph{Numerics and benchmarking:\label{subsec: Numerics}}

The particular numerical methods employed in the solution of equations~\eqref{eq: BoltzHydro}
and \eqref{eq: BoltzEvalue} are detailed in reference~\citep{Boyle2015thesis}.
Here a non-uniform energy grid is employed, that is dense at lower
energies to capture the variations near the low-energy thresholds
of the inelastic processes considered throughout this work. The other
explicit changes to the methods of~\citep{Boyle2015thesis} used
here are described in Appendix~\ref{sec: Appendix Numerical}. Systematic
benchmarking of the theory and numerical solution has been performed.

\subsection{Transport coefficients\label{subsec: TransCo}}

Knowledge of the full phase-space distribution function $f$ allows
for the calculation of all macroscopic quantities describing the electron
swarm. The distribution function $\phi_{l}(z,U,t)$, that is the solution
to equation~\eqref{eq: BoltzHydro}, allows the calculation of the
time-of-flight transport coefficients. The coefficients used here
include the mean energy $\langle\varepsilon\rangle$, flux $\left(W\right)$
and bulk $\left(W_{B}\right)$ drift velocities, the nett rate coefficient
$\left(R_{\textrm{net}}\right)$ summed over all reactive collision
frequencies $\nu_{0}^{R}(U)$, and the bulk longitudinal diffusion
coefficient $D_{B,L}$, and are defined as: 
\begin{eqnarray}
\langle\varepsilon\rangle & = & 2\pi q\left(\frac{2q}{m_{e}}\right)^{3/2}\int U^{3/2}F_{0}(U,t)dU,\nonumber \\
W_{B} & = & \frac{2\pi}{3}\left(\frac{2q}{m_{e}}\right)^{2}\int UF_{1}(U,t)dU\nonumber \\
 &  & -2\pi\left(\frac{2q}{m_{e}}\right)^{3/2}\int U^{1/2}J_{0}^{R}\left(F_{0}^{(L)}(U,t)\right)dU,\label{eq: TransCo}
\end{eqnarray}
\begin{eqnarray}
R_{\textrm{net}} & = & \sum_{R}2\pi\left(\frac{2q}{m_{e}}\right)^{3/2}\int U^{1/2}\nu_{0}^{R}(U)F_{0}(U,t)dU,\nonumber \\
D_{B,L} & = & \frac{2\pi}{3}\left(\frac{2q}{m_{e}}\right)^{2}\int UF_{1}^{(L)}(U,t)dU\nonumber \\
 &  & -2\pi\left(\frac{2q}{m_{e}}\right)^{3/2}\int U^{1/2}J_{0}^{R}\Bigg(\frac{1}{\sqrt{3}}\Bigg[F_{0}^{(2T)}(U,t)\nonumber \\
 &  & -\sqrt{2}F_{0}^{(2L)}(U,t)\Bigg]\Bigg)dU.\label{eq: TransCo1}
\end{eqnarray}
In the absence of non-conservative collisions, the bulk transport
coefficients reduce to the flux coefficients, $W$ and $D_{L}$, represented
by the first term in each of the bulk coefficient definitions.

An important self-consistency/accuracy check for an accurate solution
are the rates of energy and momentum exchange, where the gain from
the advective terms (the external electric field and time rate of
change components) and loss due to collisions must be balanced. Calculation
of energy and momentum-transfer rates due to individual cross-sections
allows assessment of the contribution of not only each collision type,
but separation into inelastic and superelastic channels. For the time-of-flight
experiment, the power exchange due to the advective terms is given
by:
\begin{eqnarray*}
P_{\textrm{adv}} & = & qE_{z}W+R_{\textrm{net}}\langle\varepsilon\rangle,
\end{eqnarray*}
while the power exchange due to each collisional process is given
by:
\begin{eqnarray*}
P_{\textrm{coll}} & = & 2q\pi\left(\frac{2q}{m_{e}}\right)^{3/2}\int_{0}^{\infty}U^{3/2}J_{0}(F_{0})dU.
\end{eqnarray*}

For the steady-state Townsend configuration, denoted by the subscript
$SST$, the mean energy and drift velocity are given by~\citep{Robson1991c,Dujko2008b}:
\begin{eqnarray*}
\left\langle \epsilon\right\rangle _{SST} & = & 2\pi\left(\frac{2q}{m_{e}}\right)^{3/2}\int U^{3/2}\psi_{0}(U,t)dU,\\
 & = & \mathbf{\varepsilon}+k_{1}\gamma+\dots,\\
W_{SST} & = & \frac{2\pi}{3}\left(\frac{2q}{m_{e}}\right)^{2}\int U\psi_{1}(U,t)dU,\\
 & = & W-k_{1}D_{L}+\dots,
\end{eqnarray*}
where $\varepsilon$ is the spatially averaged mean energy, $k_{1}$
is the  Townsend ionisation coefficient, $\gamma=\gamma\hat{\mathbf{E}}$
is the gradient energy parameter~\citep{White1995a}, and $D_{L}$
is the flux longitudinal diffusion coefficient given by the first
term in the bulk coefficient definition. In the absence of non-conservative
collisions, the $SST$ coefficients reduce to the flux coefficients.
The Townsend ionisation coefficient, calculated directly from the
solution to equation~\eqref{eq: BoltzEvalue}, can also be related
to the bulk hydrodynamic time-of-flight coefficients, when spatial
gradients are weak, through the nett reaction rate~\citep{Robson1991c,Dujko2008b}:
\begin{eqnarray}
R_{\textrm{net}} & = & k_{1}W_{B}-k_{1}^{2}D_{B,L}+\dots,\label{eq: TOF_Townsend}\\
 & = & \sum_{R}2\pi\left(\frac{2q}{m_{e}}\right)^{3/2}\!\!\!\int\!\!U^{1/2}\nu_{0}^{R}(U)\psi_{0}(U,t)dU,
\end{eqnarray}
where the summation is over all of the reactive processes $R$, like
ionisation and attachment, and the expression for the Townsend ionisation
coefficient to second order becomes $k_{1}=\frac{W_{B}}{2D_{B,L}}\pm\sqrt{\left(\frac{W_{B}}{2D_{B,L}}\right)^{2}-\frac{R_{\textrm{net}}}{D_{B,L}}}$.
The experimentally measured Townsend ionisation coefficient is related
to the calculated coefficient through $k_{1}=\alpha_{T}/n_{0}$.

\subsection{Monte Carlo technique\label{subsec: Monte-Carlo-Technique}}

We have implemented a standard swarm Monte-Carlo sampling code. The
code uses the null-collision method~\citep{Skullerud1968} along
with temperature included via appropriate modifications to the total
cross-section and resolution of collisions~\citep{RistPetr2012}.
Measurements are made through the `box sampling' style~\citep{Dujko2008},
where an integral over the quantities to be measured is performed
between each collision and binned into time bins. Hence a time-specific
measurement refers to an average of that quantity during the time
bin. To ensure we have considered a large enough simulation time to
have reached steady-state, we consider a sufficiently fine time grid
to allow a fit of the quantities to the empirical form of: $x(t)=x_{S}+e^{-\lambda_{x}(t-T/2)}\delta x$,
where $x_{S}$ is the steady state value for quantity $x$ and $T$
is the final time of the simulation. This definition allows us to
give $\delta x$ the meaning of a deviation from steady-state at the
half-way point of the simulation. The condition, $\delta x/x_{S}<10^{-4}$
is enforced, and we then average over the latter half of the simulation
to build up the statistics for the Monte-Carlo results.

We estimate the error in these results by the standard error of the
averaged simulations at different times. We have also ensured that
the autocorrelation between consecutive points is minimal.

The Monte-Carlo code has been tested against many benchmarks including
pure elastic models of hard sphere and Maxwell models~\citep{Boyle2015thesis},
argon measurements~\citep{Boyle2015Argon}, the inelastic and anisotropic
models of Reid~\citep{Reid1979,Boyle2015thesis}, the ionisation
models of Ness and Robson~\citep{NessRob1986,Boyle2015thesis}, and
inclusion of a static structure factor~\citep{Tattersall2015}.

As part of the tests to be performed in section~\ref{sec:Results-and-Discussion},
we require a different temperature for the elastic and inelastic processes.
We have implemented this by considering a mixed system of two species.
The first species possess only an elastic process, with a gas temperature
given by the elastic temperature. The second species possesses only
an inelastic process, with the ground and excited populations given
by the inelastic temperature. When the elastic and inelastic temperatures
coincide, this is equivalent to a simulation of a single species with
both processes.

\subsection{Experimental technique\label{subsec: Experimental-technique}}

The fully automated pulsed Townsend experiment, used to measure the
drift velocity and the Townsend ionisation coefficient for electrons
in gaseous $\textrm{N}_{2}$, has been described in detail previously~\citep{Hernandez2002,Basurto2013,deUrq2014}
and so the experiment is only briefly summarised here. The total displacement
current of the electrons, and their ionic products, that drift through
the parallel plate capacitor under a homogeneous electric field is
measured and separated into a fast component due to the electrons,
and a second part due to the slower ions. 

The initial swarm of electrons is generated from the cathode by an
incident 3~ns duration, UV (355~nm) laser pulse, and the electrons
and ions formed by reactions with the neutrals drift to their respective
electrodes under the action of a highly homogeneous electric field
$E$, produced by a very stable voltage in the range 0.2-5~kV, according
to the density-normalised field $E/n_{0}$ selected and the density
of the gas in the discharge vessel. The electrons and ions drift through
the capacitor with a fixed drift distance of 3.1~cm ($\pm$0.025~mm),
between an aluminium cathode and a non-magnetic stainless steel anode,
each 12~cm in diameter. The electrons are detected with a low-noise,
40~MHz amplifier with a transimpedance of $10^{5}$~V/A. The measurements
presented here were performed over the temperature range 293-300~K,
measured with a precision of $\pm$0.5~K, and with a pressure range
of 0.5-30~Torr, as monitored with an absolute pressure capacitance
transducer with 0.15\% uncertainty. The commercial grade sample of
N$_{2}$ used here from \foreignlanguage{american}{Praxair} had a
stated purity of \foreignlanguage{american}{99.995}\%. The overall
uncertainty in the measurements of the electron drift velocity was
2.2\%, and 7.4 to 9.4\% for the Townsend ionisation coefficient.

\section{Results and Discussion\label{sec:Results-and-Discussion}}

\subsection{NDC --- a model cross-section study\label{subsec:Benchmark-model}}

A limitation of the existing collision benchmark models we use is
in testing/verifying the inclusion of superelastic processes in the
inelastic channel. In the absence of superelastic processes, thermal
temperatures cannot be achieved, so a simple model system, verified
by an independent Monte Carlo method, allows us to confirm that our
solution methods are well representing physical processes.

In the pursuit of clear criteria for the existence or prediction of
NDC, a number of model cross-sections have been proposed (see for
example~\citep{Petrovic1984}). Many of these models could be adapted
to account for the inclusion of superelastic processes. The model
considered in this work, however, was chosen to illustrate the damping
effect of superelastic populations on NDC at room temperature, similar
to the behaviour of electrons in molecular nitrogen. For collisions
with neutrals with a mass $m_{0}=28\,\textrm{amu},$ at 0~K, 77~K,
and 293~K the transport coefficients have been calculated for the
model elastic and excitation cross-sections (in atomic units):
\begin{eqnarray}
\sigma_{m}^{\textrm{elas}} & = & A+BU,\nonumber \\
\sigma_{0}^{\textrm{inel}} & = & \begin{cases}
0 & U\leq0.002\,\textrm{eV},\\
A & U>0.002\,\textrm{eV},
\end{cases}\label{eq: ModelXsect}
\end{eqnarray}
where $A=1\text{\AA}^{2}$ and $B=5\text{\AA}^{2}/\textrm{eV}$.

\begin{figure}
\includegraphics[width=0.9\columnwidth]{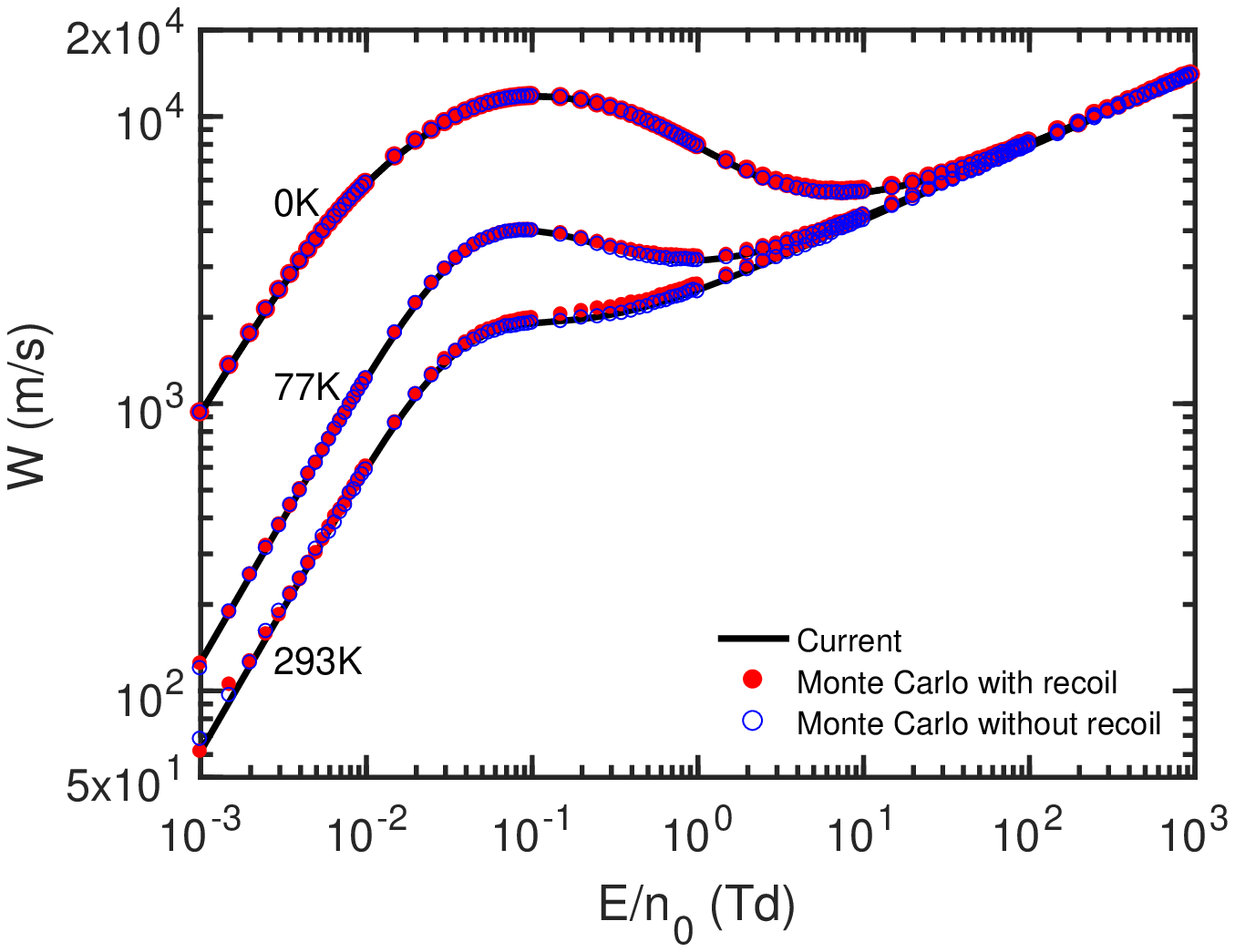}

\includegraphics[width=0.9\columnwidth]{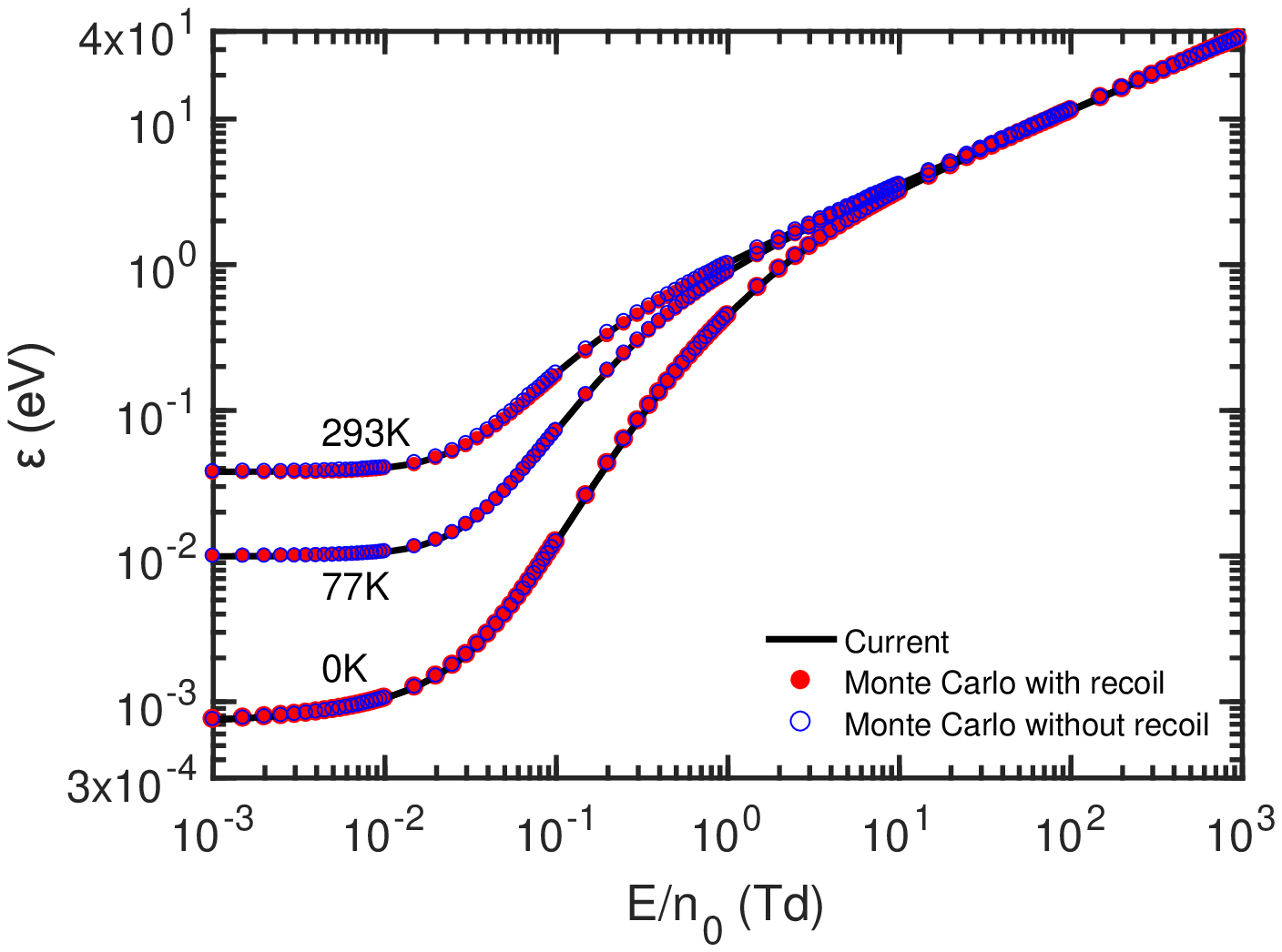}\foreignlanguage{american}{\protect}

\caption{The calculated flux drift velocity (upper) and mean energy (lower)
of a swarm of electrons whose collisional behaviour is described by
the proposed model cross-section set. The lines represent values calculated
at various temperatures using the multi-term kinetic theory, while
the symbols show the values calculated using an independent Monte
Carlo solution method. The solid red symbols represent the Monte Carlo
calculations treating inelastic collisions exactly, while the open
blue symbols correspond to Monte Carlo calculations with recoil of
the neutral particle during inelastic collisions neglected.}
\label{fig: Model_WEn}
\end{figure}

Figure~\ref{fig: Model_WEn} shows the drift velocity and mean energy
calculated using the multi-term Boltzmann equation solution and the
independent Monte Carlo code, as a function of the reduced electric
field $E/n_{0}$ in units of the Townsend (1 Td = $10^{-21}\textrm{Vm}^{2}$).
The agreement between the Monte Carlo and Boltzmann solutions is better
than 2\% for the drift velocities and the mean energies at 0~K, and
with less than a 4\% variation in the drift velocity and a 2\% variation
in the mean energy at 77~K. However, this increases to 6\% for both
the drift velocity and mean energy at 293~K.

To address the discrepancy between the transport coefficients calculated
using our Boltzmann and Monte Carlo solutions, we consider the neglect
of recoil in the inelastic channel in our solution (and similar solutions
of the Boltzmann equation, for example the recent work of Ridenti
\emph{et al.~}\citep{Ridenti2015} in the continuous energy loss
approximation). Unlike elastic collisions, that are represented to
first order in the mass ratio to take into account the thermal motion
and recoil of the neutral particle during an elastic collision, recoil
of the neutral particle during inelastic collisions is neglected in
most of the existing Boltzmann equation solutions, and in our solution.
This assumption has been considered previously in White \emph{et al.}~\citep{WhiteMorrMas2002},
using the integral form of the inelastic collision operator that does
not restrict collisional representation to zeroth-order. The transport
coefficients, calculated for electron impact on H$_{2}$ using a multi-term
solution over the range 0.1--10~Td, differed by less than 0.1\%
between no recoil inelastic collisions and the converged collision
description. In their work, the lowest excitation channel in H$_{2}$
is the $0\rightarrow2$ rotational excitation with a threshold of
44~meV, while for our model system the energy loss threshold is 2~meV,
much closer to the first order mass ratio $\frac{m_{e}}{m_{e}+m_{0}}=0.02$~meV
for the model system.

\begin{figure}
\includegraphics[width=0.9\columnwidth]{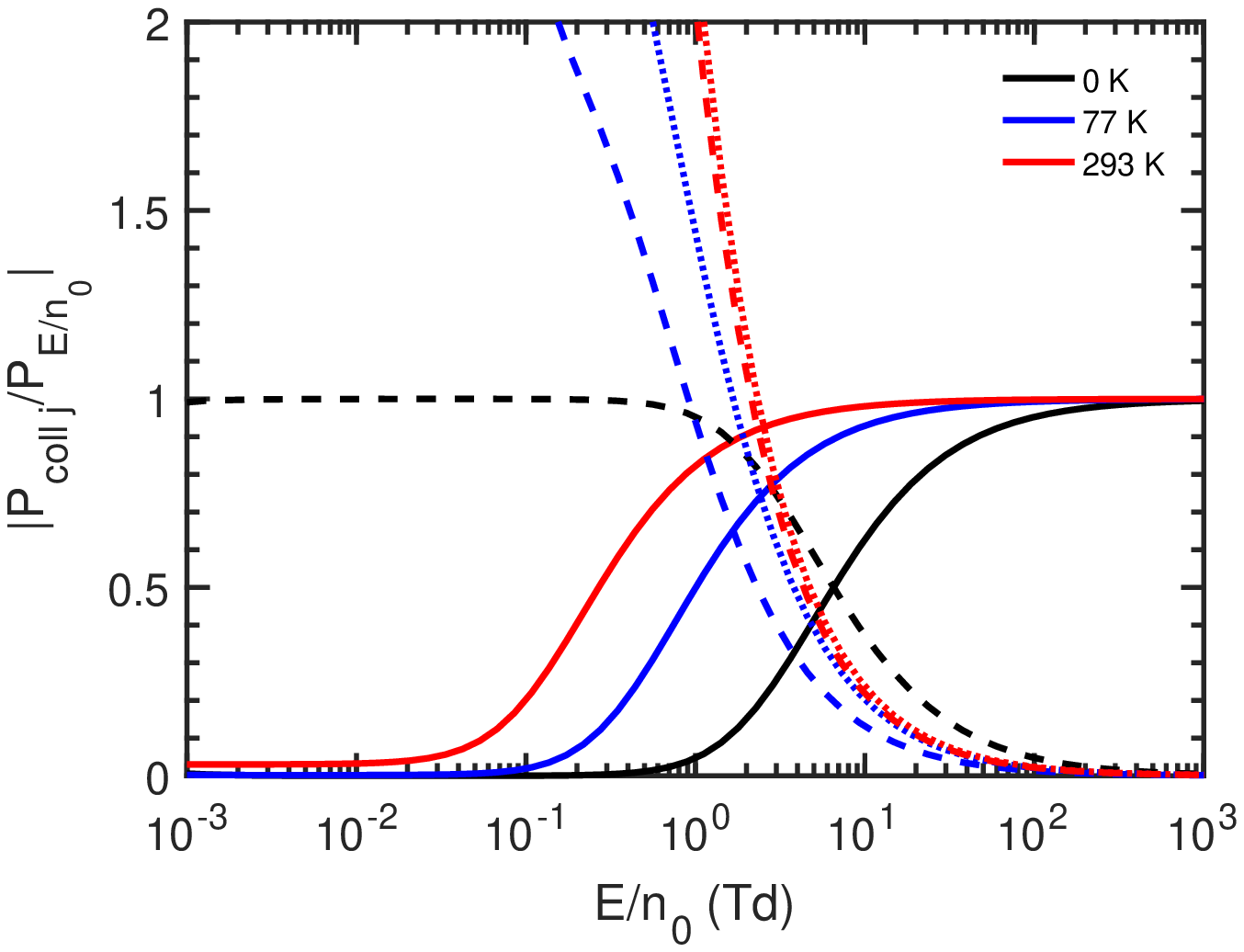}

\includegraphics[width=0.9\columnwidth]{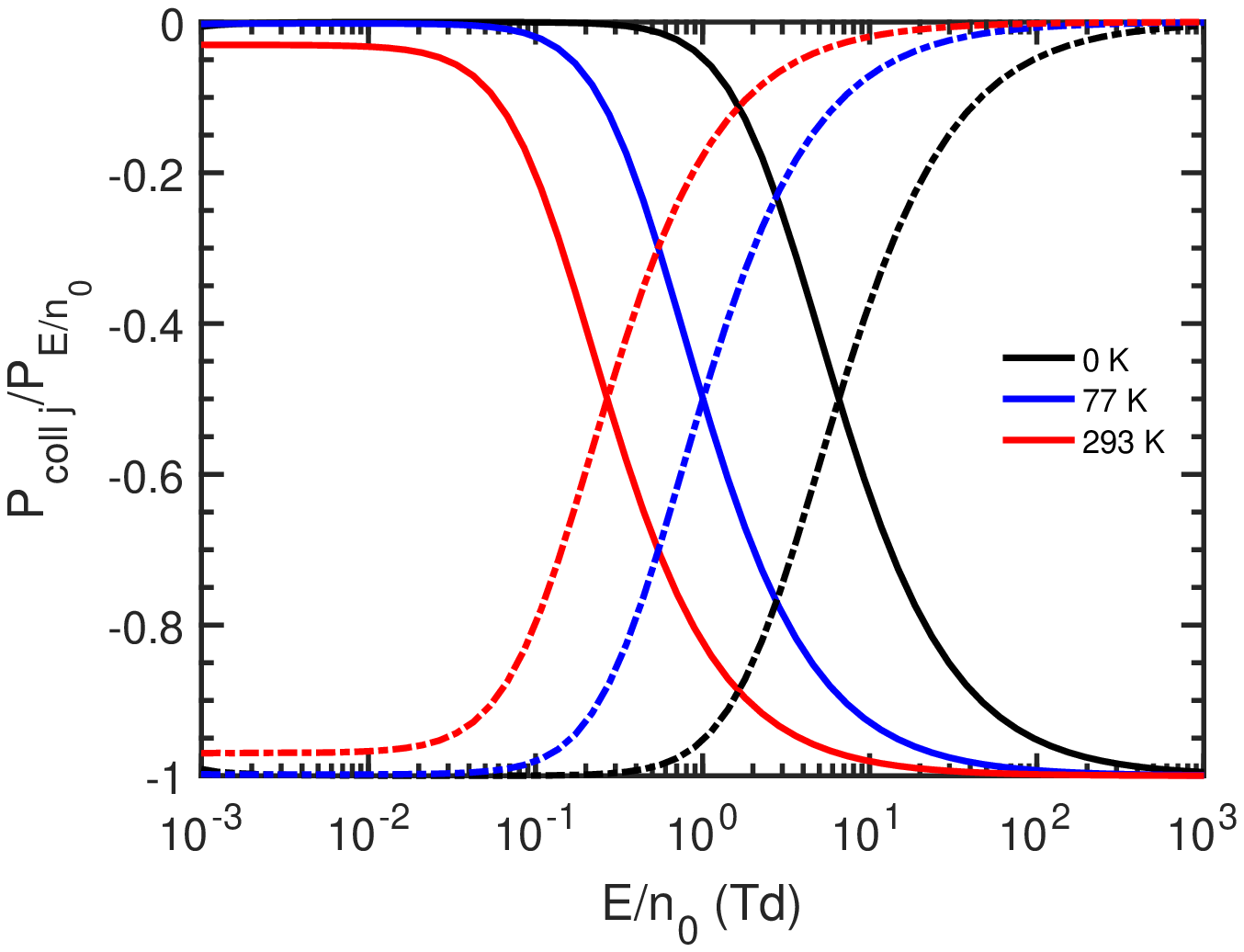}\foreignlanguage{american}{\protect}

\caption{The energy transfer rates of each of the collision channels $j$ of
the model cross-section, as a fraction of the power input from the
external reduced electric field, at different temperatures, as a function
of the reduced electric field. (Upper) The solid lines show the fractional
power transfer rates of the elastic cross-section, the dashed lines
represent the ground-state inelastic process, and the dotted lines
represent the gain in energy due to the superelastic process only.
(Lower) The nett energy transfer rates, as a fraction of the reduced
electric field, where the solid lines represent the elastic channel,
and the dot-dashed lines correspond to the nett power transfer from
the inelastic channel.}
\label{fig: Model_P}
\end{figure}

To include the thermal motion of the neutrals during inelastic collisions
in the Frost-Phelps differential finite difference collision operator,
requires an extension that is outside the scope of this study. However,
we still desire quantification of the effect on the transport coefficients.
In Monte Carlo simulations the collisions are treated exactly, so
the Monte Carlo technique described in section~\ref{subsec: Monte-Carlo-Technique}
was used to assess the effect of recoil in the low-threshold channel
of interest here, as shown in figure~\ref{fig: Model_WEn}. The effect
of truncation of the mass ratio for inelastic collisions on our calculations
is most prevalent at reduced electric fields between 0.1~Td and 10~Td,
where the nett energy transfer due to elastic collision is increasing
relative to the energy transfer due to inelastic collisions, as shown
in figure~\ref{fig: Model_P}. Here, the difference between the complete
and approximate collision descriptions in the Monte Carlo calculations
is greatest at 293~K, where the drift velocity and mean energy both
differ by up to 7\%. At 77~K, the differences are up to 4\% and 3\%
between the drift velocity and mean energy, respectively, and at 0~K
the drift velocity and mean energy differ by 1.7\% and 1.3\%, respectively,
between the two collision representations.

When recoil of the neutral particle during inelastic collisions in
our Monte Carlo simulation is neglected by artificially increasing
the neutral mass for inelastic collisions only, to replicate the differential
finite difference form of the inelastic collision operator utilised
here, the difference between our Monte Carlo and Boltzmann calculations
reduces to 0.9\% and 1.3\% for the drift velocity and mean energy,
respectively, at 0~K, 1.4\% and 0.8\% between the drift velocity
and mean energy at 77~K, and generally below 2.8\% (increasing to
4.7\% at low fields with statistical noise) and 1.1\% between the
drift velocity and mean energy at 293~K, respectively. 

The presence of NDC is anticorrelated with the presence of superelastic
collisions, highlighting the damping effect of the de-excitation process
on the presence of NDC. When properly included through detailed balance
for inelastic collisions, a smaller ratio of neutrals in the ground-state,
caused by an increasing temperature, increases the mean energy of
the swarm and decreases the drift velocity. The energy transfer profiles
given in figure~\ref{fig: Model_P} show the lower nett inelastic
energy transfer rate with increasing temperature, increasing the mean
energy which samples higher-energy regions of the elastic cross-section,
resulting in a reduced average velocity of the swarm. NDC ceases when
the collisional energy transfer is dominated by the elastic process.
At higher temperatures, the increased fraction of neutrals in excited-state
populations reduces the nett power transfer due to inelastic collisions,
as shown by the superelastic contribution to the energy transfer in
figure~\ref{fig: Model_P}. As a direct result of the superelastic
population, the range of NDC is reduced and the transition to elastic-collision
dominated energy transfer occurs at lower reduced electric fields
for increasing temperatures.


\subsubsection{Criterion for NDC\label{subsec: Criteria-for-NDC}}

Using momentum-transfer theory, Robson's~\citep{Robs84} criterion
for the presence of NDC uses the energy variation of the ratio of
the elastic to inelastic energy transfer. For a monotonically increasing
elastic collision frequency and open inelastic channels, the mean
energy increases with increasing reduced electric field, slowly as
the inelastic collisions take energy from system, so that the drift
velocity increases with field, as illustrated in figure~\ref{fig: Model_WEn}.
As the inelastic collisions become less important relative to the
elastic collisions, the mean energy of the swarm increases at a greater
rate, to sample the higher elastic collision frequency, causing the
drift velocity to begin to decrease with increasing field. The criterion
derived for the appearance of NDC by Robson~\citep{Robs84}, at a
given mean energy $\varepsilon$, is given by $1+\frac{\partial\Omega}{\partial\varepsilon}<0$
where $\Omega$ represents the ratio of the total inelastic to elastic
energy transfer and is given by:
\begin{eqnarray}
\Omega & \equiv & \frac{\sum_{j}U_{th}^{j}\left\{ \langle\nu_{0}^{\textrm{inel},j}(jk;\varepsilon)\rangle-\langle\nu_{0}^{\textrm{ inel},j}(kj;\varepsilon)\rangle\right\} }{\frac{2m_{e}}{m_{0}}\langle\nu_{m}^{\textrm{elas}}\left(\varepsilon\right)\rangle},\label{eq: RR_Omega}
\end{eqnarray}
where the total inelastic energy transfer is taken as the sum over
all inelastic channels $j$ with associated thresholds $U_{th}^{j}$.

We note that the NDC criteria of Robson~\citep{Robs84} and Petrovi{\'c}
\emph{et al.}~\citep{Petrovic1984} differ due to a different expression
for the energy balance, where the latter omit energy transfer due
to elastic collisions, although this is sufficient for the systems
considered in that work. 

\begin{figure}
\includegraphics[width=0.9\columnwidth]{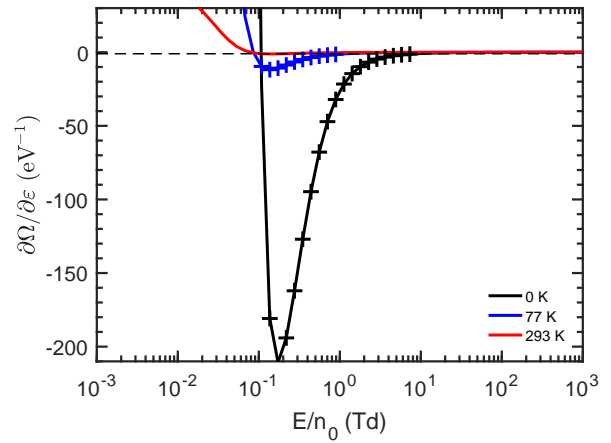}\foreignlanguage{american}{\protect}

\caption{{Rate of change of $\Omega$ with mean energy as a function of the
reduced electric field, for a swarm of electrons whose collisional
behaviour is described by our model cross-sections. The solid lines
show $\partial\Omega/\partial\varepsilon$ at varying temperatures,
where below $-1$ (indicated by the dashed
horizontal line) NDC is predicted by Robson's
criterion~\citep{Robs84}. The symbols indicate where NDC is present
in our calculated values. We note that the prediction of NDC at 0~K
appears to occur earlier than its appearance in the calculated data;
however, this is only due to the rapid decrease
in the derivative around 0.1~Td.}}

\label{fig: SuperElas_OmegaField}
\end{figure}

The criterion proposed by Robson~\citep{Robs84} is a very good predictor
for NDC for the model cross-section considered in this study, as shown
in figure~\ref{fig: SuperElas_OmegaField}. The NDC region for 0~K
and 77~K ceases when the energy transfer rate due to elastic collisions
is greater than the nett energy transfer rate due to the inelastic
process, as predicted. For the calculation of $\Omega$ at 293~K,
for reduced electric fields between 0.1~Td and 0.2~Td, the derivative
is $\approx-1$ and the presence of NDC is only weakly predicted,
but does not occur in our calculations.

\subsubsection{Temperature dependence and detailed balance}

To illustrate the physical dependence on the inclusion of superelastic
collisions, in this subsection we consider two unphysical modifications
to our model system that address the effect of temperature from each
of the scattering channels separately. The two models considered incorporate
different temperatures for the background gas through elastic collisions
and excited state populations. 

\begin{figure}
\includegraphics[width=0.9\columnwidth]{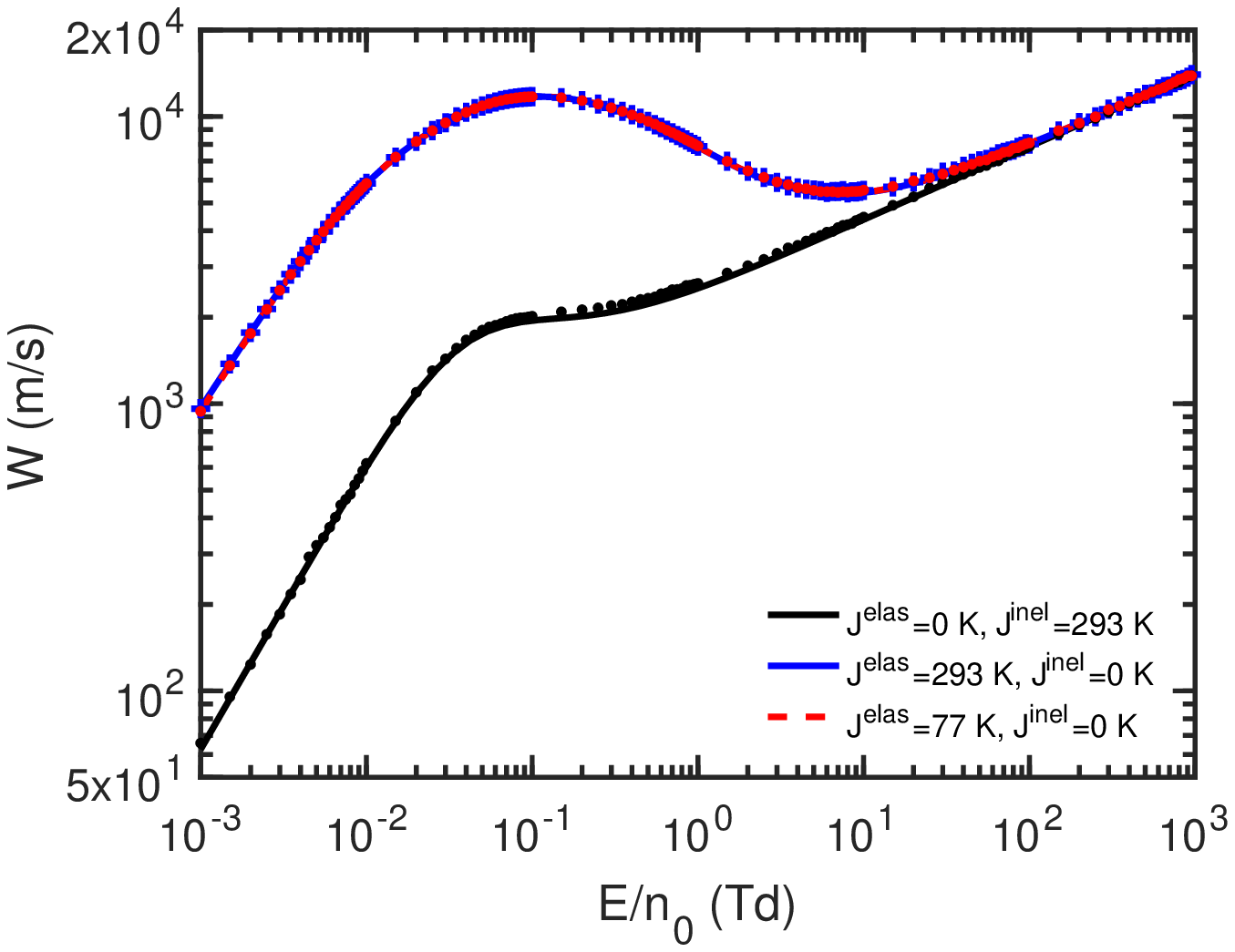}

\includegraphics[width=0.9\columnwidth]{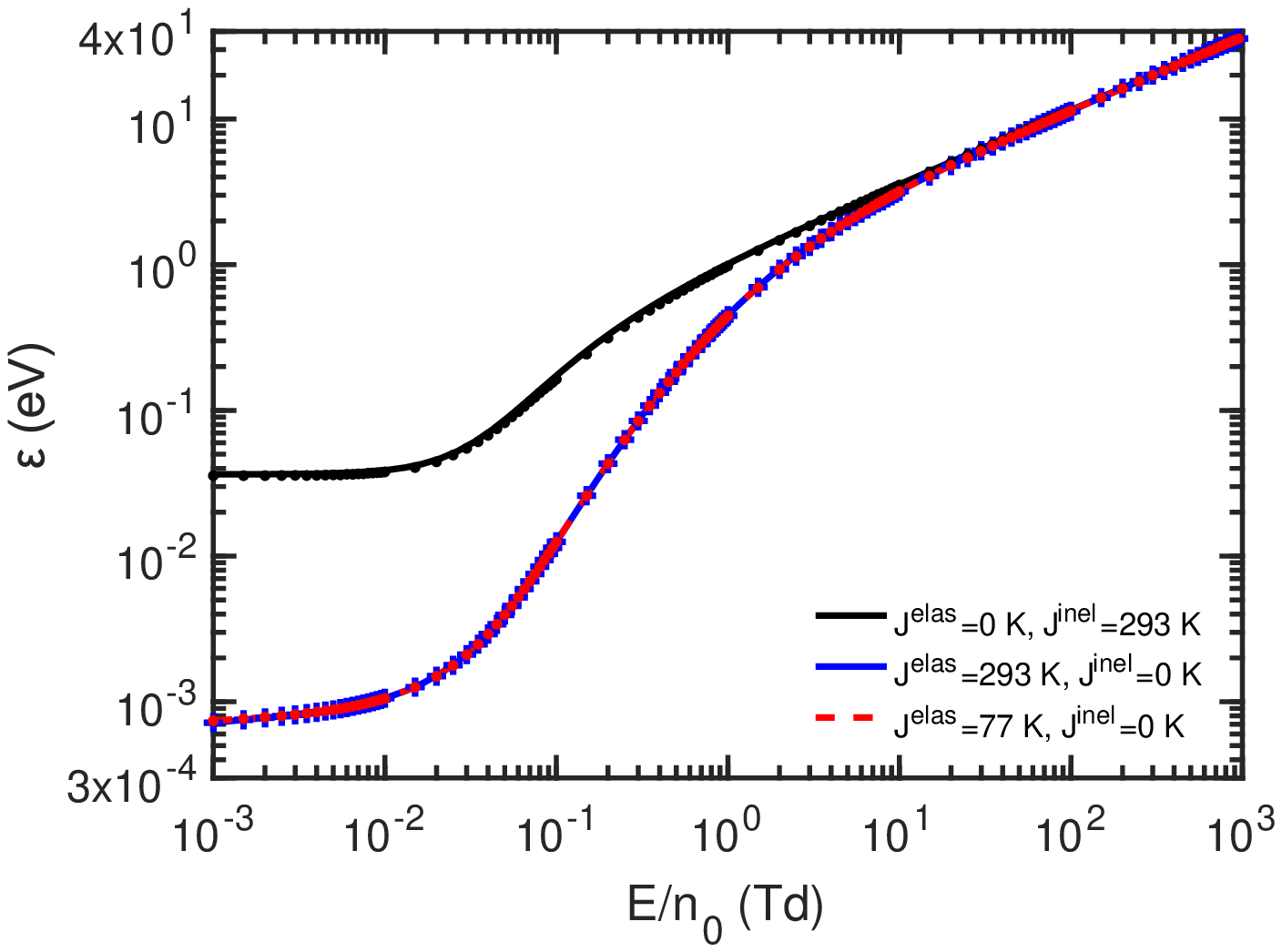}\foreignlanguage{american}{\protect}

\caption{The calculated flux drift velocity (upper) and mean energy (lower)
of a swarm of electrons whose collisional behaviour is described by
the proposed model cross-section set. The lines represent values calculated
for the various models using the multi-term kinetic theory solution,
while the symbols correspond to values calculated using an independent
Monte Carlo method, where the same colour denotes the same model.
See text for details of the model notation.}
\label{fig: Model_driftMeanEn-1}
\end{figure}

\begin{figure}
\includegraphics[width=0.9\columnwidth]{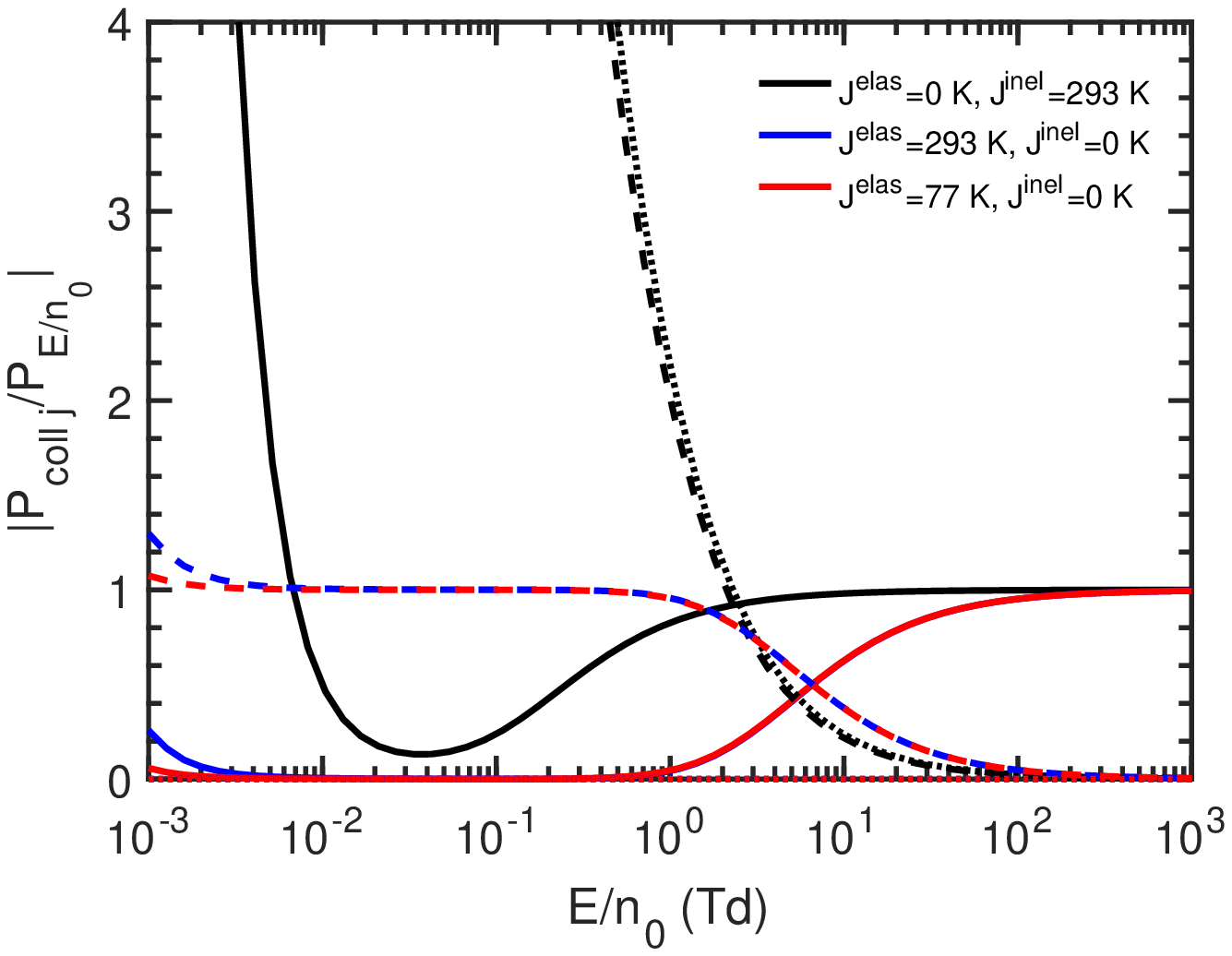}

\includegraphics[width=0.9\columnwidth]{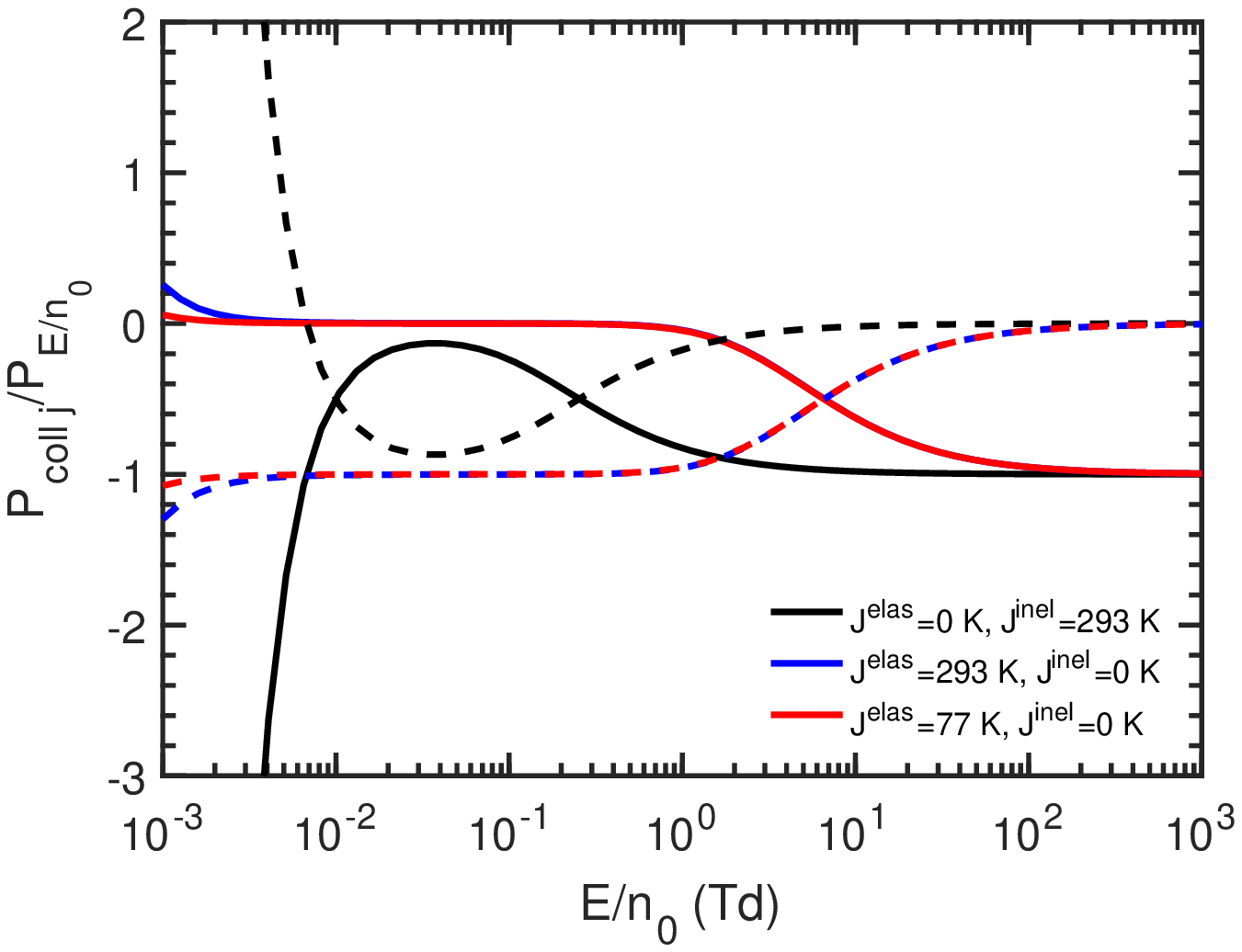}\foreignlanguage{american}{\protect}

\caption{The energy transfer rates for each of the collision channels of the
model cross-section, with varying reduced electric fields, as a fraction
of the power input from the external reduced electric field. (Upper)
The solid lines show the transfer rates of the elastic cross-section,
the dashed lines represent the ground-state inelastic process, and
the dotted lines represent the gain in energy due to the superelastic
process only. (Lower) The energy transfer rates for the elastic (solid
lines) and nett inelastic (dashed lines) collisional channels. For
the two models where superelastic collisions are neglected, the positive
values indicate energy is being lost in the elastic channel at very
low reduced electric fields to balance the power gain from the external
electric field. See text for details of the model notation.}
\label{fig: Model_EnMTTransferRates-2}
\end{figure}

\begin{figure}
\includegraphics[width=0.9\columnwidth]{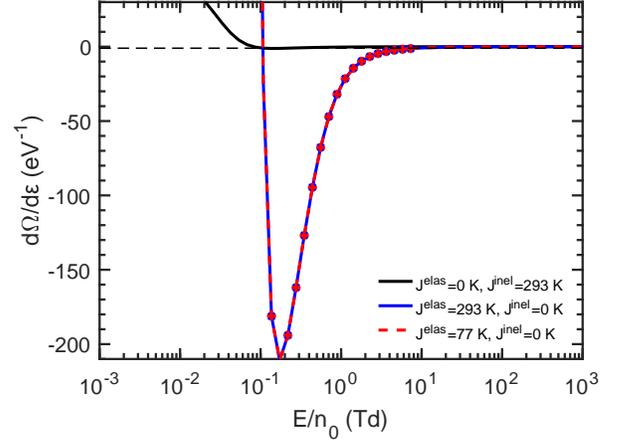}\foreignlanguage{american}{\protect}

\caption{{Rate of change of $\Omega$ with mean energy as a function of reduced
electric field for a swarm of electrons whose collisional behaviour
is described by our model cross-section. The solid lines show $\partial\Omega/\partial\varepsilon$
for the various models, where below $-1$
(indicated by the dashed horizontal line)
NDC is predicted by Robson's criterion~\citep{Robs84}. The symbols
indicate where NDC is present in our calculated values. We note the
prediction of NDC in models with all neutrals in the ground state
(blue and red) appears to occur earlier
than its appearance in the calculated data,
this appearance is only due to the rapid decrease in the derivative
around 0.1~Td. See text for details of the model notation.}}

\label{fig: Model_OmegaField_2}
\end{figure}

The first model includes temperature dependence of the excited state
population, but considers elastic collisions with stationary neutrals,
equivalent to a temperature of 0~K in the elastic collision operator,
with the notation $J^{\textrm{elas}}=0$~K, $J^{\textrm{inel}}=293$~K
in the following figures. The second model involves elastic collisions
with non-stationary neutrals, at 293~K and 77~K, but inelastic collisions
from ground- to excited-states only, with the notation $J^{\textrm{elas}}=77/293$~K,
$J^{\textrm{inel}}=0$~K in figures~\ref{fig: Model_driftMeanEn-1}--\ref{fig: Model_OmegaField_2}. 

Figure~\ref{fig: Model_driftMeanEn-1} displays the transport coefficients
from our Boltzmann solution, with a zeroth order mass ratio representation
in the inelastic collision integral, and compares them with our independent
Monte Carlo solution, with inelastic collisions treated exactly, for
each of these models. For the model $J^{\textrm{elas}}=0$~K, $J^{\textrm{inel}}=293$~K,
differences of less than 6\% and 5.4\% were found between the drift
velocity and mean energy, respectively. For the two models with elastic
collisions taken to be with non-stationary neutral particles and inelastic
collisions between ground-state neutrals only, we find differences
of less than 2.5\% and 1.4\% for elastic collisions at 293~K, and
2\% and 1.3\% when elastic collisions are taken at 77~K, between
the drift velocity and mean energy, respectively.

For properly included superelastic collisions, but when temperature
is not included in the elastic channel, the mean energies approach
the appropriate thermal value of $\approx\frac{3}{2}k_{B}T_{0}$ with
decreasing reduced electric field. At 293~K a difference of around
3.6\% between the model and standard calculation is observed, generally
decreasing with increasing reduced electric field strength. We find
that the drift velocity calculations lie very close to, but just above,
the standard model calculations at 293~K, by at most 2.2\%. While
those differences are not as significant as some of the others discussed
in this study, these calculations do illustrate the importance of
detailed balance in swarm calculations for achieving correct thermal
distributions.

A more dramatic difference is observed when temperature effects are
taken into account for the elastic collisions, but not in the inelastic
channel. Here, the calculated drift velocity and mean energy approach
the 0~K calculations due to the dominance of the inelastic channel
at low reduced electric fields, as can be seen in the energy transfer
profiles given in figure~\ref{fig: Model_EnMTTransferRates-2}. The
difference between these models and the 0~K profiles shows the explicit
contribution of the temperature term in the elastic collision operator.
The variation from the standard temperature treatment profiles is
large, and results in an overestimate of the drift velocity below
30~Td and the presence of an NDC region that is larger than that
at 77~K and absent from the 293~K calculations. 

When temperature effects are included through the elastic collision
operator and the inelastic ground-state density, but detailed balance
is not achieved due to the neglect of superelastic collisions altogether,
we observe a dramatic effect on the transport coefficients, given
the large contribution of the de-excitation process to the energy
transfer. Although not shown, the resulting drift velocity and mean
energy profiles lie between the 0~K and 293~K results, as is expected
with less energy lost in the inelastic channel than the 0~K simulation,
but a greater nett energy loss in the inelastic channel than the 293~K
simulation, where the de-excitation collisions contribute to energy
gained by the electron swarm. These model systems highlight the necessity
of detailed balance in collisional processes when modelling real gaseous
systems in the low energy regime.

For these non-physical models, which disregard thermal effects in
the elastic, inelastic and superelastic channels, NDC is present in
our results for the two cases where the de-excitation process is removed.
Similar to the results shown above, the presence of superelastic collisions
increases the mean energy of the electron swarm, increasing the drift
velocity monotonically. As demonstrated by the earlier results, Robson's
criterion for the presence of NDC gives a very accurate prediction
based only on a knowledge of the energy transfer rates. Illustrated
in figure~\ref{fig: Model_OmegaField_2} is the rate of change with
energy of the ratio of the inelastic to elastic energy transfer rate,
$\frac{d\Omega}{d\varepsilon}$, where the rapid decrease in this
rate to below -1 at around 0.1~Td corresponds to the start of the
NDC region for both of the models with no superelastic processes.
Regardless of the temperature of the model system, in the absence
of the de-excitation process the ratio of the energy transfers in
equation~\eqref{eq: RR_Omega} decreases more rapidly with energy
than it would otherwise, resulting in NDC until the elastic energy
transfer rate starts to dominate.


\subsubsection{Approximation effects: Anisotropy in the inelastic channel and higher-order
superelastic terms\label{subsec: Model_approximations}}

In this study we are interested in the effect of anisotropic scattering
in low-threshold processes like rotational excitations. The impact
on the calculated transport coefficients are explored in this subsection,
alongside our assessment of other assumptions, including the two-term
approximation and those associated with the inclusion of superelastic
collisions.

\begin{figure}
\includegraphics[width=0.9\columnwidth]{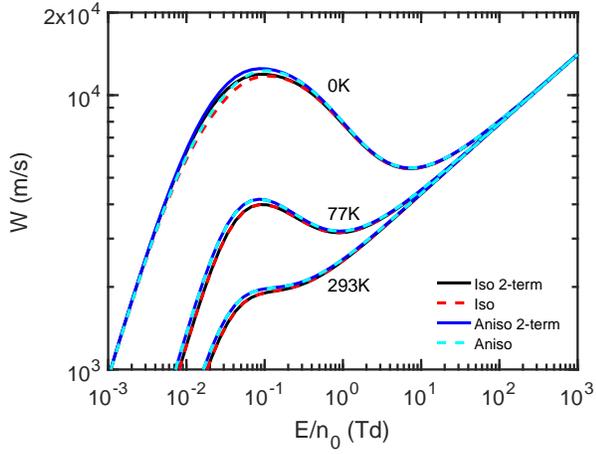}

\selectlanguage{american}%
\protect

\caption{Variation of the drift velocity with reduced electric field for the
scattering model detailed in equations~\eqref{eq: ModelXsect} and
\eqref{eq: Model_xsect_aniso}, at different temperatures. Results
are shown for a two-term approximation (labelled 2-term), multi-term
calculation for the isotropic model (labelled Iso) and anisotropic
scattering in the inelastic channel (labelled Aniso).}

\label{fig: Model_anisotropic}
\end{figure}

The effect of anisotropic scattering in the inelastic channel has
been recently investigated in Janssen \emph{et al.}~\citep{Janssen2016},
for an excitation scattering channel of (simplified) argon with a
threshold at 11.828~eV. We note this energy threshold is much higher
than the lowest inelastic thresholds of molecules like N$_{2}$. For
the low threshold processes considered here, to quantify the effects
of the anisotropic terms in the inelastic operator, given in equations~\eqref{eq: InelasCollOp}
and \ref{eq: InelasCollOp2}, we introduce an angular scattering component
for our model excitation in order to emulate the forward-peaked nature
of rotational excitation. Using a forward scattering model where the
differential inelastic cross-section is $\sigma^{\textrm{inel}}(U,\chi)\propto\cos\frac{\chi}{2}$,
the inelastic partial cross-sections are given by:
\begin{align}
\sigma_{l\geq1}^{\textrm{inel}} & =\frac{1}{5}\sigma_{0}^{\textrm{inel}}.\label{eq: Model_xsect_aniso}
\end{align}
To test explicitly the assumption of isotropic scattering in the inelastic
channel, here we modify only the inelastic momentum-transfer, leaving
the elastic momentum-transfer cross-section fixed, as has been considered
previously by Reid~\citep{Reid1979} and Phelps and Pitchford~\citep{PhelpsPitch1985_aniso},
for example. Note that this does not fix the total momentum-transfer
cross-section. Our calculated drift velocities for the anisotropic
model combining equations~\eqref{eq: ModelXsect} and \eqref{eq: Model_xsect_aniso}
are shown in figure~\ref{fig: Model_anisotropic}. For the various
temperatures considered in this work, the effects of anisotropy in
the inelastic channel are greatest where momentum exchange is dominated
by the inelastic channel. At 0~K, this difference occurs over the
range 0.02 to 0.2~Td with a variation of less than 5\% in the drift
velocity and less than 9\% in the mean energy of the swarm (not plotted).
For the 77~K and 293~K simulations, the maximum difference occurs
at the lowest reduced electric fields, where the momentum exchanged
during superelastic collisions increases the total momentum exchanged
in the inelastic channel. At 77~K the drift velocity and mean energy
differ by 11\% and 6.5\% respectively, and 10\% and 4\% at 293~K,
respectively, decreasing with increasing reduced electric field for
both temperatures.

The validity of using a two-term approximation has been discussed
previously (e.g., in~\citep{White2003a,PhelpsPitch1985_aniso}),
and we briefly consider the effects of that assumption on our model
calculations here. Figure~\ref{fig: Model_anisotropic} shows our
calculated drift velocity for isotropic and anisotropic scattering
using the model cross-sections of equations~\eqref{eq: ModelXsect}
and \eqref{eq: Model_xsect_aniso}. For both the isotropic and anisotropic
models, the difference between the two-term and multi-term results
is greatest at 0~K, with a difference of up to 8\% for reduced electric
fields below 0.1~Td in the drift velocity, and 5\% in the mean energy.
While at 77~K and 293~K, for both models, the differences between
these transport coefficients reduces to below 0.3\%.

\begin{figure}
\includegraphics[width=0.9\columnwidth]{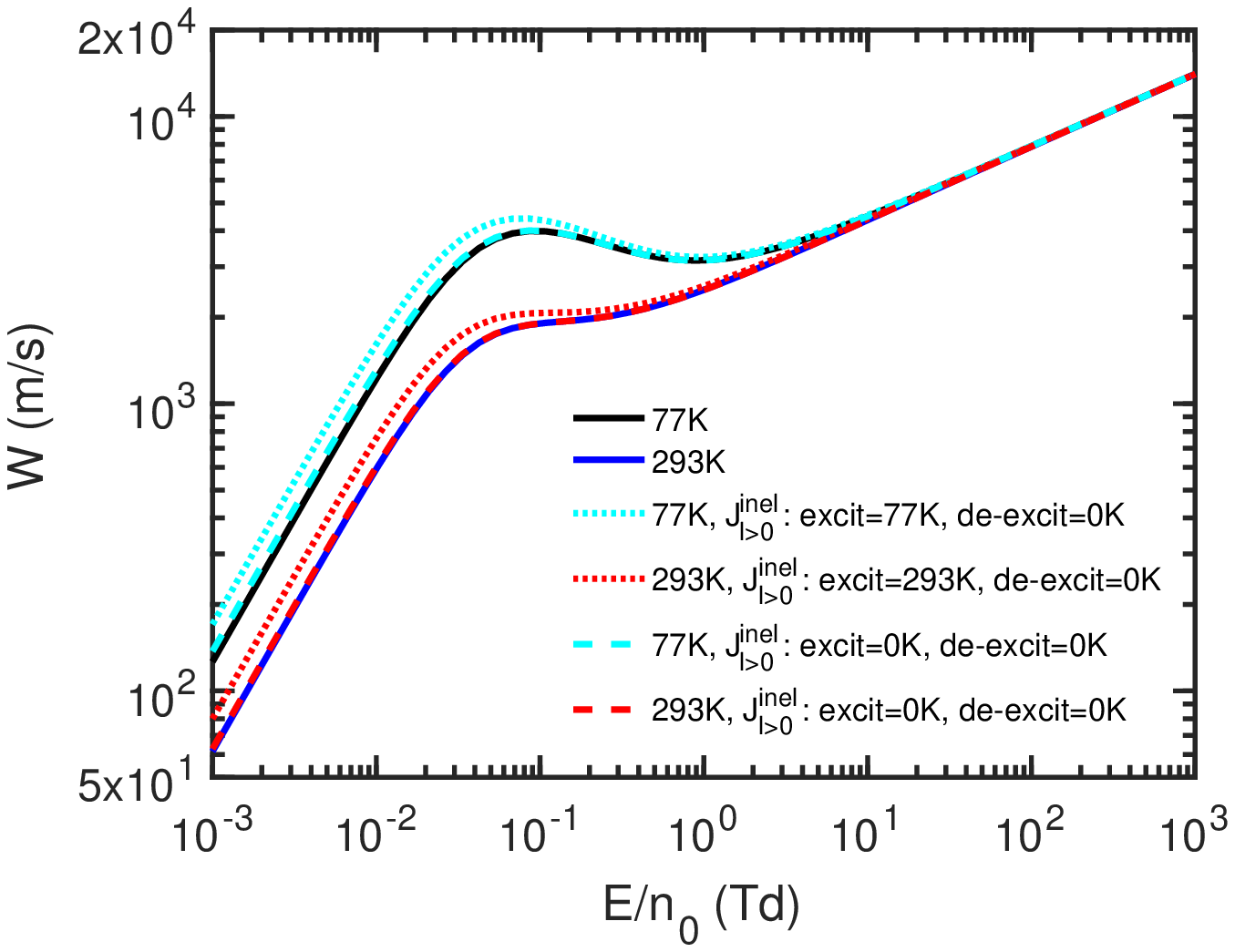}

\includegraphics[width=0.9\columnwidth]{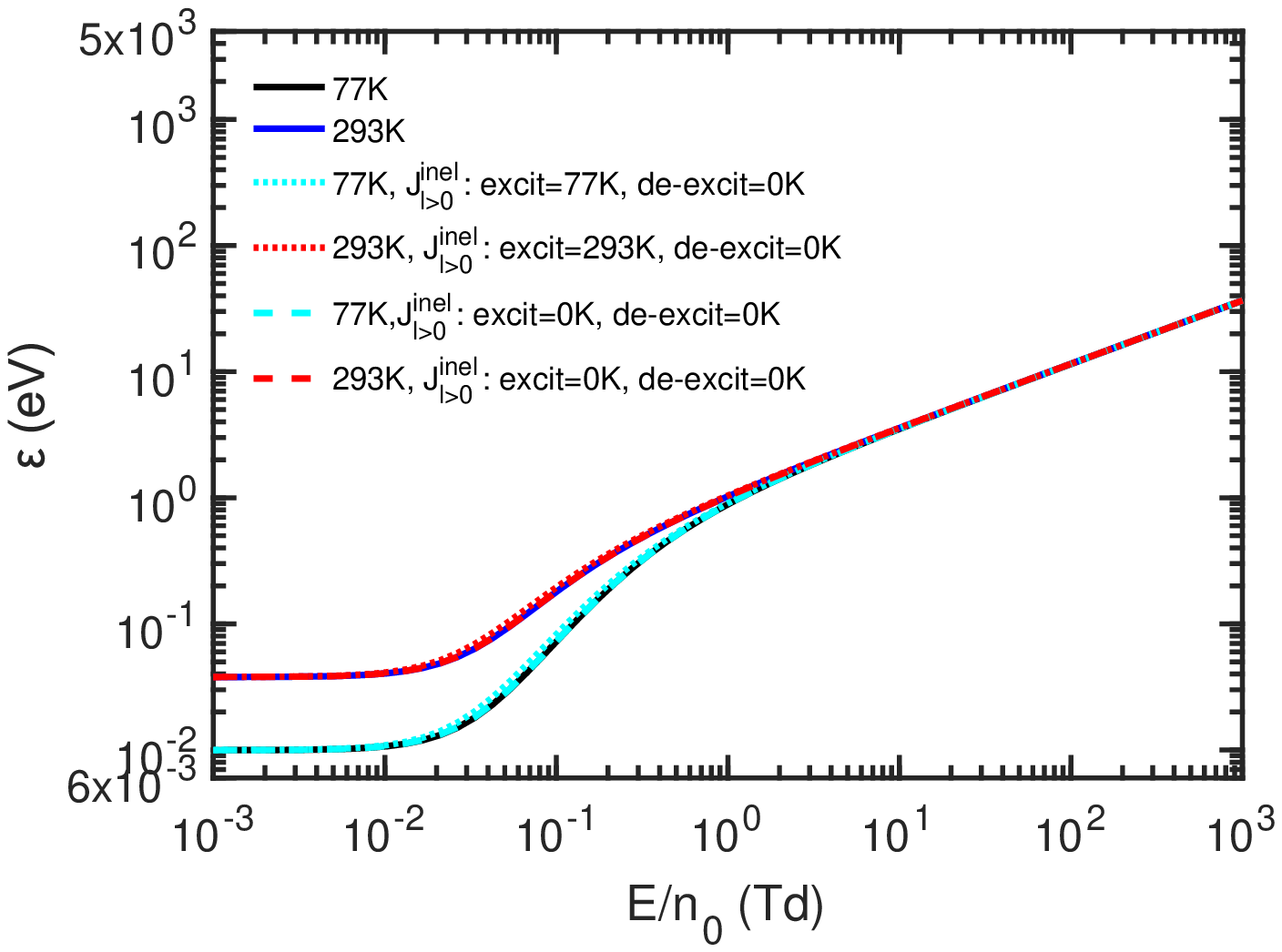}\foreignlanguage{american}{\protect}

\caption{Variation of the drift velocity (upper) and mean energy (lower) with
reduced electric field for the isotropic scattering model~\eqref{eq: ModelXsect},
at 77~K and 293~K when superelastic collisions are included or neglected
in the $J_{l\protect\geq1}^{\textrm{inel}}$ collision term. For each
calculation temperature is included through the elastic collision
operator, and $l=0$ inelastic collision terms. Compared with our
standard calculations (solid lines) is a model with the ground-state
population in the $l\protect\geq1$ inelastic collision operator calculated
according to the neutral temperature and de-excitations neglected
(dotted lines), and another with all neutrals in the ground state
in the $J_{l\protect\geq1}^{\textrm{inel}}$ equation (dashed lines).}
\label{fig: Model_HigherOrderSuper}
\end{figure}

The final assumption we wish to address is that while de-excitation
is considered in the $l=0$ equation of the inelastic operator, it
is sometimes neglected in the $l\geq1$ equations. For our isotropic
model detailed in equation~\eqref{eq: ModelXsect}, we have removed
superelastic collisions in the $l\geq1$ channels in two ways. First
we consider the proportion of particles in the ground-state calculated
according to the neutral temperature, as is included through the $l=0$
equation, but simply turn off the de-excitation channel, denoted by
the notation $J_{1}^{\textrm{inel}}:\,\textrm{excit}=77\,\textrm{K},\,\textrm{de-excit}=0\,\textrm{K}$,
for example, in figure~\ref{fig: Model_HigherOrderSuper}. The differences
between the calculated drift velocity and mean energy, when compared
with our standard treatment, are up to 33\% and 16\% at 77~K and
30\% and 10\% at 293~K, respectively. We also consider neglecting
higher-order superelastic terms by setting all neutral particles in
the ground state, with the notation $J_{1}^{\textrm{inel}}:\,\textrm{excit}=0\,\textrm{K},\,\textrm{de-excit}=0\,\textrm{K}$,
and find much smaller differences as the total number of excitation
collisions remains constant, with differences of less than 9\% and
1\% in the drift velocity and mean energy, respectively, at 77~K,
and 2\% and 0.3\% at 293~K. The magnitude of these differences decreases
with increasing reduced electric field, influenced by the relative
strength of the two cross-sections, and the dominance of the elastic
cross-section above 1~Td. We have also repeated these calculations
using our anisotropic model detailed in equations~\eqref{eq: ModelXsect}
and \eqref{eq: Model_xsect_aniso}, and find that the differences
are very similar in magnitude, as is expected by the small difference
($\frac{1}{5}\sigma_{0}$) between the momentum-transfer cross-section
that enters the inelastic collision operator (equation~\eqref{eq: InelasCollOp2})
for isotropic scattering, and the anisotropic form.

For each of the cases tested, our calculations demonstrate that significant
differences can appear when various approximations are made, or detailed
balance is neglected. Of particular interest in this study is anisotropic
scattering in the low-threshold inelastic channel for our model system,
where differences up to 11\% in the transport coefficients were calculated
from the inelastically-isotropic model.

\subsection{Electron transport in molecular nitrogen\label{subsec:N2}}

\subsubsection{Cross-section set\label{subsec: N2_Cross-section-set}}

The set of $\textrm{N}_{2}$ cross-sections utilised throughout this
work were obtained from the v8.97 Magboltz database tabulated on LXCat~\citep{Biagi_lxcat}.
Other databases of scattering cross-sections for electron swarms in
molecular nitrogen are available (e.g., see~\citep{Pitchford2017}),
but our focus here is to illustrate the damping effects of temperature
on NDC in a real gas rather than an analysis of the self-consistency
of the different sets. The cross-section set~\citep{Biagi_lxcat}
details elastic collisions, 15 individual vibrational states, 29 electronic
state collisions and a single ionisation cross-section, with rotational
excitations to be included as in the Magboltz source code~\citep{Magboltz897},
outlined below. We also note that the elastic momentum-transfer cross-section
tabulated on LXCat is more sparse at lower energies than in the source
work of Itikawa~\citep{Itikawa2006}, so the tabulation from Itikawa
at lower energies is preferred. 

\begin{figure}
\includegraphics[width=0.9\columnwidth]{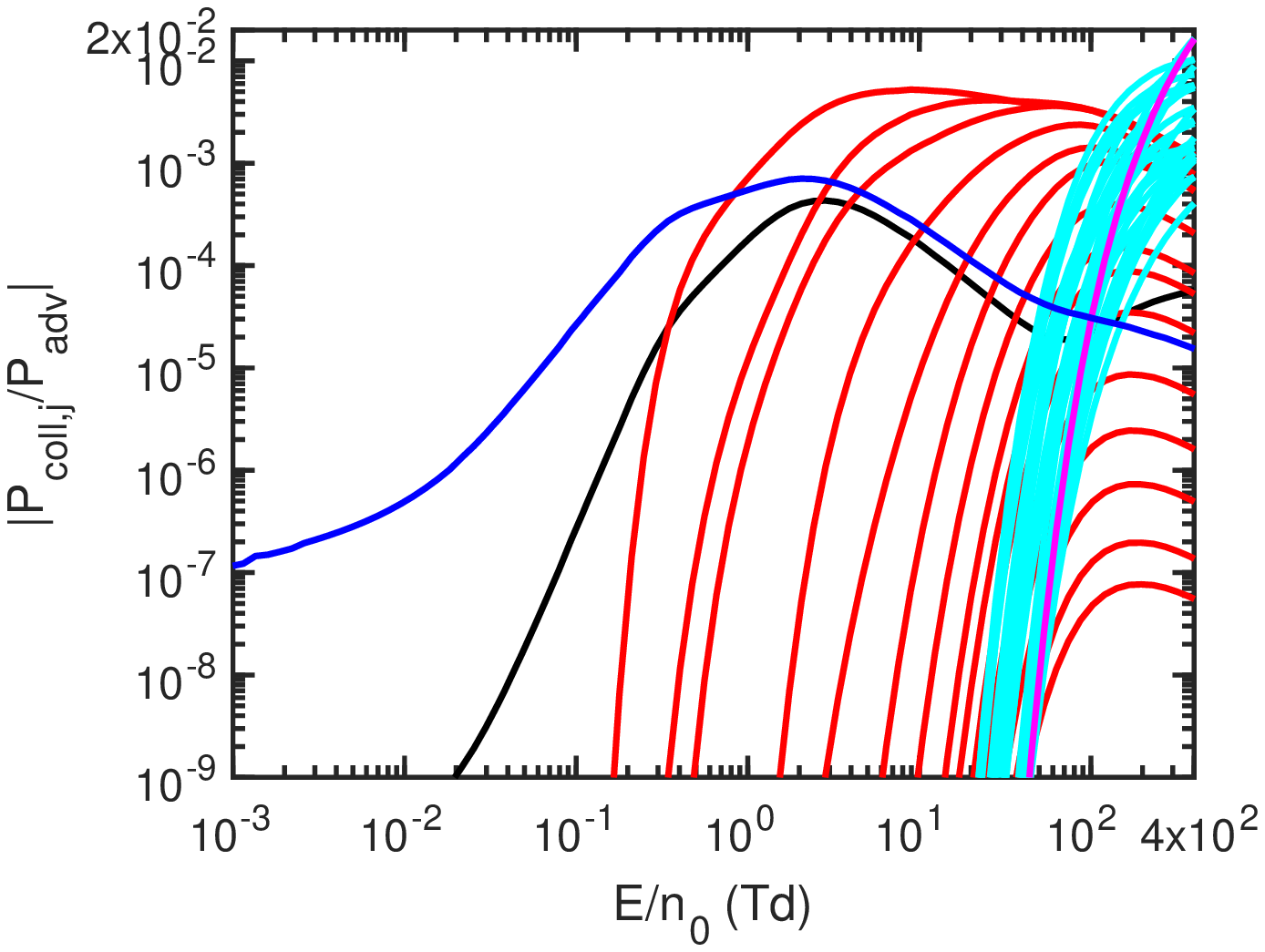}

\includegraphics[width=0.9\columnwidth]{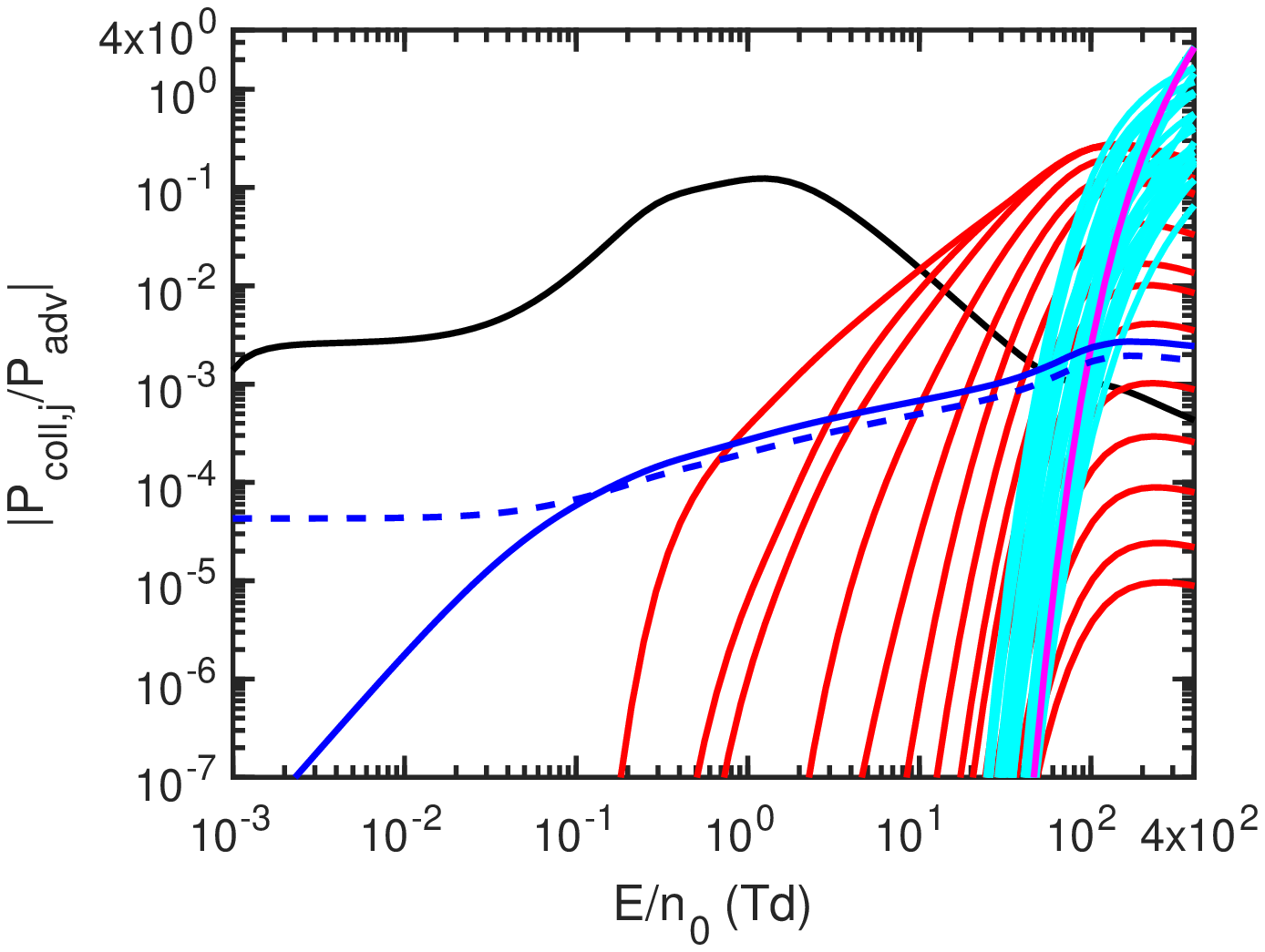}\foreignlanguage{american}{\protect}\caption{The nett power transfer rates for each collision type, as a fraction
of the advective power transfer, for electron impact on N$_{2}$ at
0~K (upper) and 77~K (lower), as a function of reduced electric
field. The solid lines represent the nett power transfer due to individual
collisional channels, while the dashed lines correspond to the explicit
contribution to power transfer due to superelastic collisions. The
coloured lines represent: black---elastic, blue---sum of rotational
excitations (grouped for the figures only), red---vibrations, cyan---electronic-state,
and magenta---ionisation.}
\label{fig: N2_EnTrf}
\end{figure}

The LXCat Biagi database at present does not include rotational cross-sections,
and although other databases do include a description of rotational
collisions, the rotations of the source work (Magboltz) are preferred
for their consistency with the elastic momentum-transfer cross-section.
The rotational excitations utilised in the Magboltz v8.97 source code~\citep{Magboltz897},
to be included with the LXCat tabulation for low field simulations,
were calculated using the Gerjuoy and Stein treatment of the atoms
as point quadrupoles with the Born approximation~\citep{GerjuoyStein1955a,GerjuoyStein1955b}.
Rotational cross-sections for the transitions $j\rightarrow j\pm2$
were considered, using the values for the quadrupole moment constant
$Q=1.045$ in units of $qa_{0}^{2}$, where $a_{0}$ is the Bohr radius,
and the rotational constant $B_{0}=2.4668\times10^{-4}$~eV~\citep{Itikawa2006,Magboltz897}.
In Magboltz, an enhancement of the cross-section magnitude in the
resonance region between 1.2~eV and 5.3~eV was explicitly included
in the rotational cross-sections. This is also included here, along
with scaling of the rotational cross-sections above 5~eV to fall
at the same rate as the elastic momentum-transfer cross-section. For
the temperatures considered in this work, sufficient convergence in
the calculated transport coefficients is obtained using 40 rotational
transition cross-sections, and we adopt no fewer here. Note that the
Born approximation, Gerjuoy and Stein, rotational cross-sections are
arguably valid up to $\sim$0.6~eV~\citep{GerjuoyStein1955a}. For
the transport coefficients considered in this work, up to 400~Td,
the rotational cross-sections are each calculated up to 400~eV. At
greater than approximately 20~Td, however, when vibrational processes
are active, the power (figure~\ref{fig: N2_EnTrf}) and momentum-transfer
rates (not shown) indicate that rotational collisions, both inelastic
and superelastic terms, are considerably less important than for the
other processes.

\begin{figure}
\includegraphics[width=0.9\columnwidth]{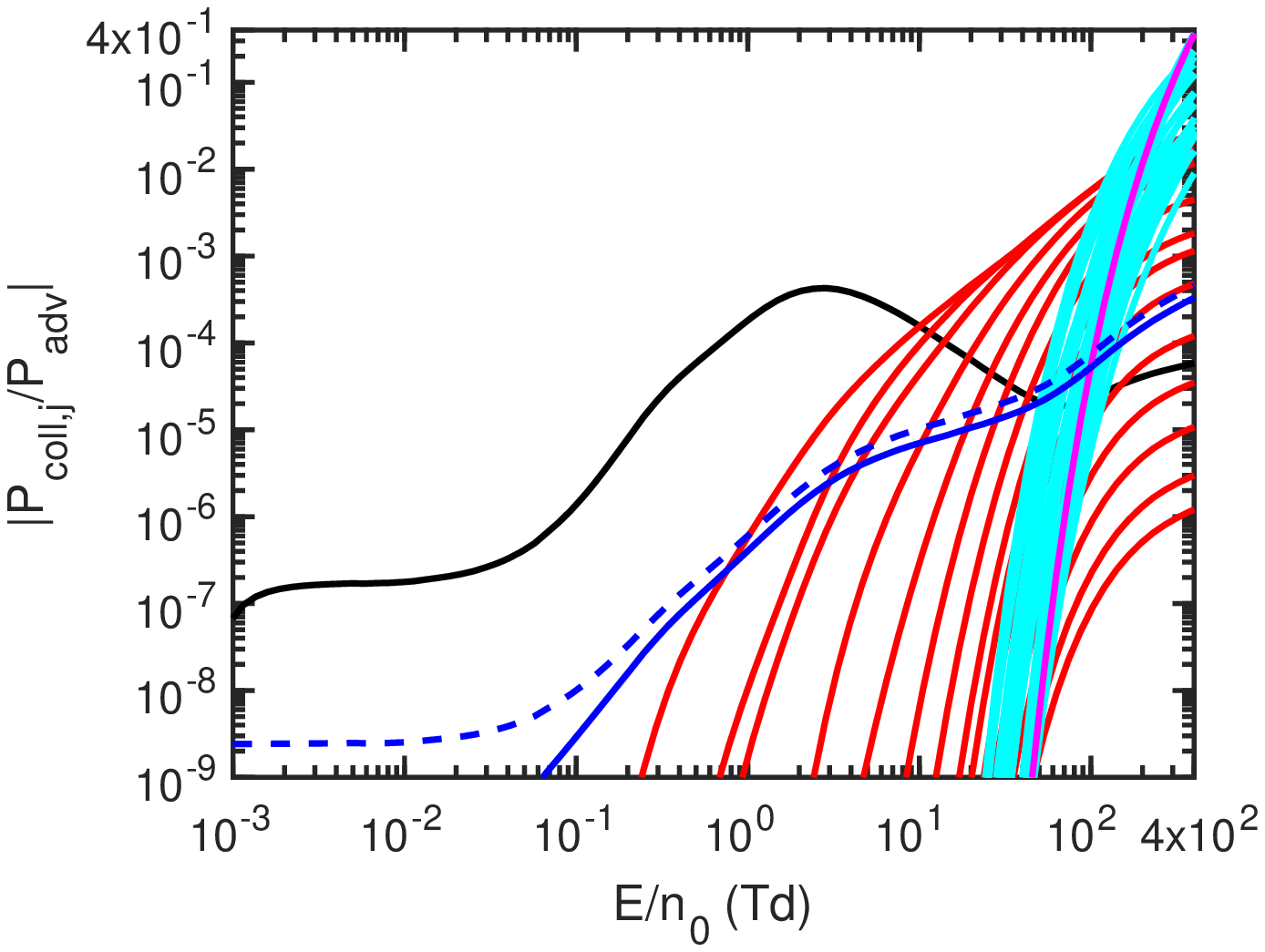}

\includegraphics[width=0.9\columnwidth]{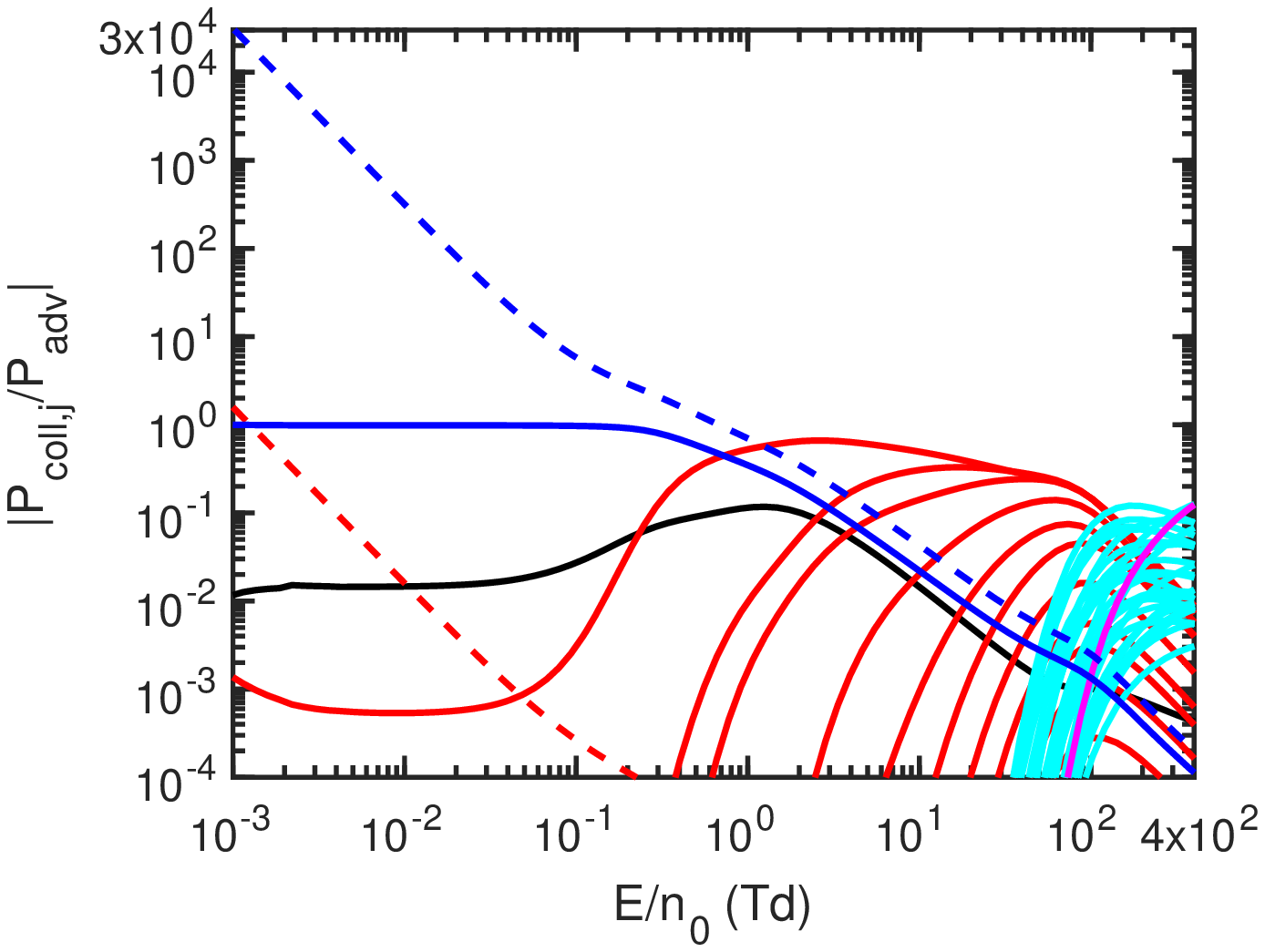}\foreignlanguage{american}{\protect}\caption{The nett power transfer rates for each collision type, as a fraction
of the advective power transfer, for electron impact on N$_{2}$ at
195~K (upper) and 293~K (lower), as a function of reduced electric
field. The solid lines represent the nett power transfer due to individual
collisional channels, while the dashed lines represent the explicit
contribution to power transfer due to superelastic collisions. The
coloured lines correspond to: black---elastic, blue---sum of rotational
excitations (grouped for the figures only), red---vibrations, cyan---electronic-state,
and magenta---ionisation.}
\label{fig: N2_EnTrf2}
\end{figure}

We also consider the energy sharing fraction between the two post-collision
electrons resulting from the ionising collisions, taken here to be
equally shared between the scattered and ejected electrons. At 360~Td,
the highest reduced electric field measured in our experiment (see
figure~\ref{fig: N2_WTI} and Appendix~\ref{sec: ExpTransCo}),
we find a difference of less than 0.4\% and 0.6\% between the drift
velocities and mean energies, respectively, from the 50\%-50\% sharing
fraction and 1\%-99\% sharing fraction. Comparing the 50\%-50\% sharing
fraction results with those for all-fractions being equiprobable,
a less than 0.02\% difference is calculated between the drift velocities,
and less than 0.3\% between the respective mean energies. The size
of these differences is not unexpected given that the power transfer
from the ionisation channel is of a similar magnitude to the (individual)
electronic state excitations at the highest reduced electric field
considered, as shown in figures~\ref{fig: N2_EnTrf} and \ref{fig: N2_EnTrf2}.

\subsubsection{Temperature dependence of the electron transport properties in N$_{2}$
--- measured and calculated\label{subsec: Temperature-dependence-N2}}

\begin{figure}[!h]
\includegraphics[width=0.9\columnwidth]{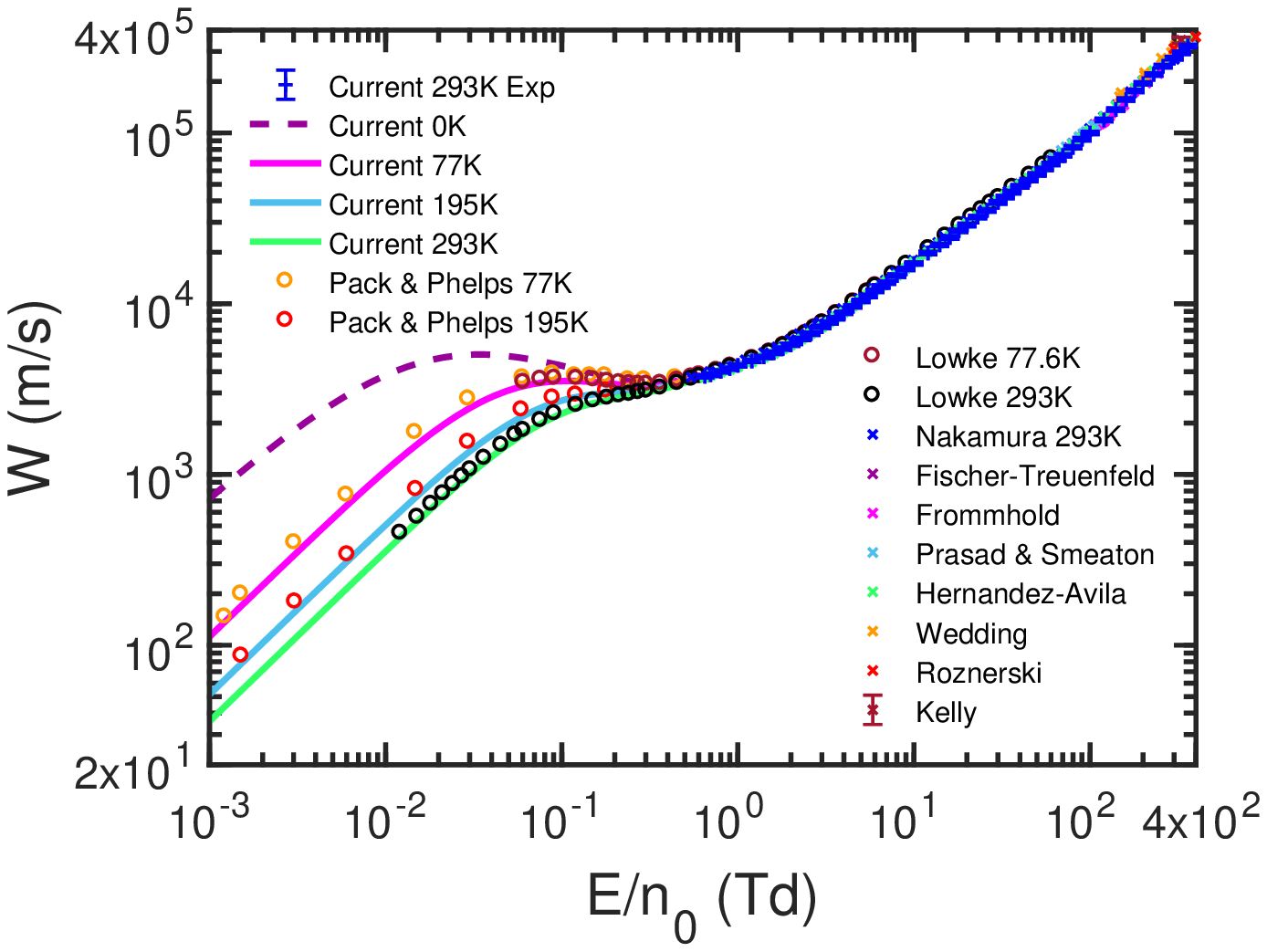}

\includegraphics[width=0.9\columnwidth]{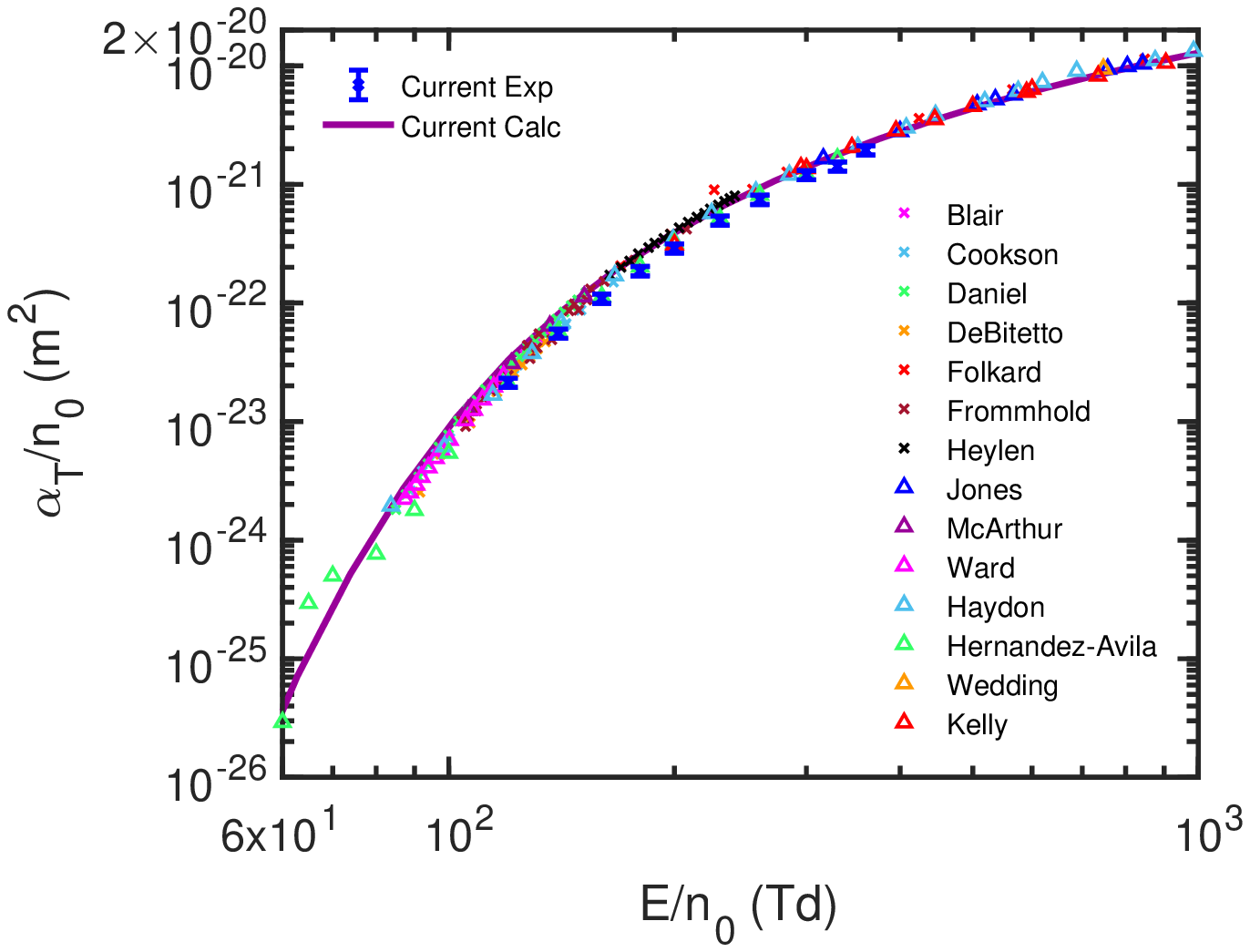}\foreignlanguage{american}{\protect}

\caption{Comparison of our experimental drift velocity (upper) and the Townsend
ionisation coefficient (lower) measured for an electron swarm in $\textrm{N}{}_{2}$
as a function of reduced electric field, with our calculated values,
and some of the other available experimental measurements at various
temperatures. Our new drift velocity measurements (blue + symbols)
at 293~K are compared with our calculations at 0~K, 77~K, 195~K,
and 293~K, depicted by the solid lines, and the experimental measurements
of Pack and Phelps at 77~K and 195~K~\citep{PackPhelps1961,PackPhelps_lx},
Lowke at 77.6~K and 293~K~\citep{Lowke1963}, Nakamura at 293~K~\citep{Nakamura1987},
Fischer-Treuenfeld~\citep{FischerT1965,FischerT_lx}, Frommhold~\citep{Frommhold1960,Frommhold_lx},
Prasad and Smeaton~\citep{Prasad1967,Prasad_lx}, Hern{\'a}ndez-{\'A}vila
\emph{et al.~}\citep{Hernandez2004}, Wedding \emph{et al.}~\citep{Wedding1985},
Roznerski~\citep{Roznerski1996,Roznerski_lx}, Kelly~\citep{Kelly1990thesis}(digitised
from Campbell \emph{et al.~}\citep{Campbell2001}). Our  Townsend
ionisation coefficient measurements (blue dots) are compared with
our calculated values (solid line) and the experimental measurements
of Bagnal and Haydon~\citep{Bagnal_lx}, Blair~\citep{Blair_lx},
Cookson \emph{et al.}~\citep{Cookson_lx}, Daniel and Harris~\citep{Daniel_lx},
DeBitetto and Fisher~\citep{DeBitettoFisher_lx}, Dutton \emph{et
al.~}\citep{Dutton_lx}, Folkard and Haydon~\citep{Folkard_lx},
Frommhold~\citep{Frommhold_lx}, Heylen~\citep{Heylen_lx}, Jones~\citep{Jones_lx},
McArthur and Tedford~\citep{McArthur_lx}, Ward~\citep{Ward_lx},
Haydon and Williams~\citep{Haydon_lx}, Hern{\'a}ndez-{\'A}vila \emph{et
al.}~\citep{Hernandez2004,Lxcat_UNAM}, Kelly~\citep{Kelly1990thesis}
(digitised from Campbell \emph{et al.}~\citep{Campbell2001}), and
Wedding \emph{et al.~}\citep{Wedding1985}.}
\label{fig: N2_WTI}
\end{figure}

In figure~\ref{fig: N2_WTI} and Appendix~\ref{sec: ExpTransCo}
we present our experimental values of the drift velocity and the Townsend
ionisation coefficient for electrons in molecular nitrogen, as a function
of the reduced electric field $E/n_{0}$ in units of the Townsend
(1 Td = $10^{-21}\textrm{Vm}^{2}$). The experimental drift velocities
measured at 293~K and over the range 0.65--360~Td are in reasonable
agreement, but tend to lie below, the other available experimental
data (by at most $\pm$10\%, with the exception of the
Wedding \emph{et al.~}\citep{Wedding1985}, Roznerski~\citep{Roznerski1996,Roznerski_lx},
and Kelly~\citep{Kelly1990thesis,Campbell2001} data at the higher
$E/n_{0}$ where that difference increases up to 19\%). Our measurements
also lie below our calculations at 293~K, with differences of less
than 2.8\% over the range of reduced electric fields measured, which
is somewhat larger the overall uncertainty of $\pm2.2\%$. 

Our experimental measurements of the Townsend ionisation coefficient
compare reasonably well with the other available experimental measurements;
within 5\% of the measurements of DeBitetto and Fisher~\citep{DeBitettoFisher_lx},
within 14\% of the Hern{\'a}ndez-{\'A}vila \emph{et al.} measurements~\citep{Hernandez2004,Lxcat_UNAM}
and those from Cookson \emph{et al.}~\citep{Cookson_lx} and
Kelly~\citep{Kelly1990thesis,Campbell2001}, and generally underestimating
the other experimental measurements shown in figure~\ref{fig: N2_WTI}
by less than 50\%, with the exception of the measurements of Haydon
and Williams~\citep{Haydon_lx}, and some of the
Wedding \emph{et al.}~\citep{Wedding1985} measurements, although
the bulk of these data lie within $\pm$5\% of our present measurements.
Comparison between our experimentally measured Townsend ionisation
coefficient and our calculations showed similar discrepancies. Namely,
our measurements are lower than our calculated values over the range
of reduced electric fields 120--360~Td, with a difference around
55\% at the lower fields where the coefficient is rapidly rising,
and decreasing to a difference of around 14\% at 360~Td. These differences
are larger than the experimental overall error bars of $\pm9\%$.
Compared with the other available measurements, our calculations tend
to overestimate the experiment below 200~Td by generally less than
60\%, decreasing with increasing field, to be within 15\% at the higher
reduced electric fields. 

Comparing our transport coefficients calculated at 293~K with our
calculations at 300~K, they show an up to 3\% variation, decreasing
below 0.1\% above 1~Td, in the drift velocity and mean energies,
and a less than 0.01\% difference in the  Townsend ionisation coefficient
above 100~Td, increasing to 1.25\% at 60~Td.

The observed discrepancies between our calculated and measured transport
coefficients are addressed in the following section.


\subsubsection{Approximation effects: Two-term approximation, drift velocity definition,
and higher order superelastic processes\label{subsec: N2_ApproximationEffects}}

Before our discussion of the effect of temperature and superelastic
populations on NDC in N$_{2}$, we digress to consider the effect
of some of the approximations associated with calculating transport
coefficients for electron swarms in N$_{2}$. This subsection details
the effect on the calculated transport coefficients of the different
definitions of the drift velocity, the two-term approximation, and
the treatment of superelastic collisions.

\begin{figure}
\includegraphics[width=0.9\columnwidth]{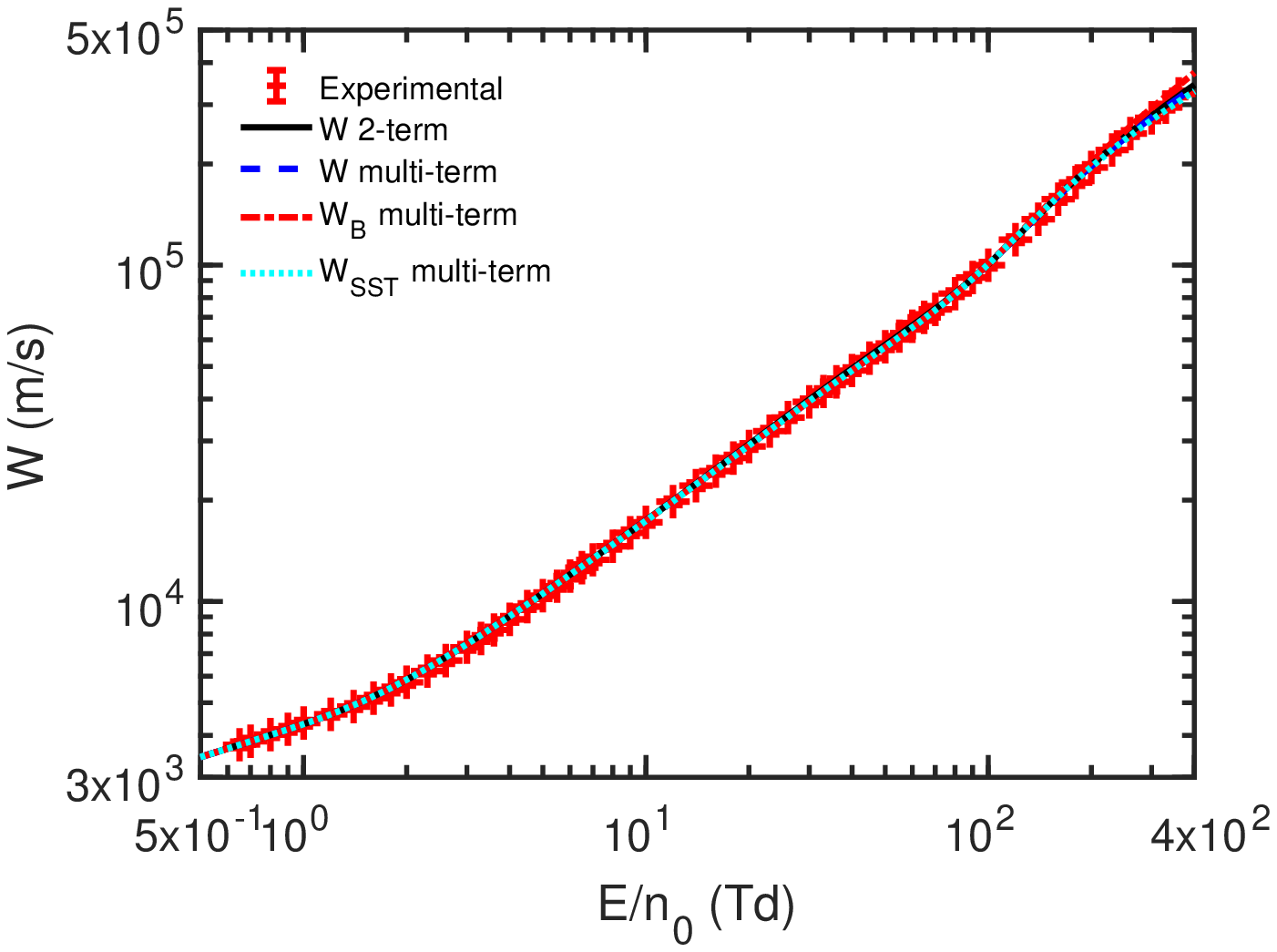}

\includegraphics[width=0.9\columnwidth]{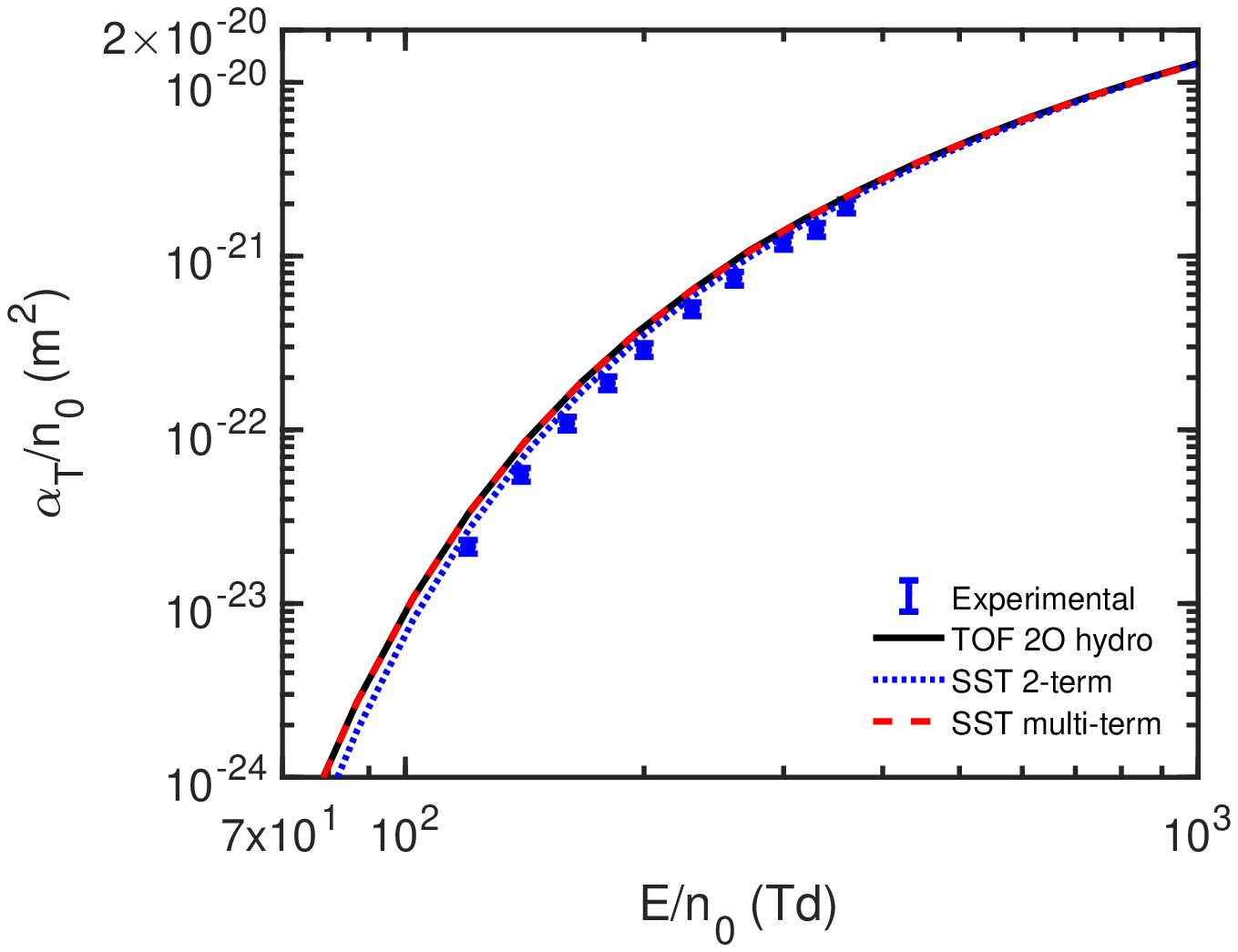}\foreignlanguage{american}{\protect}

\caption{Comparison of our experimental and calculated drift velocity and the
Townsend ionisation coefficient for electron impact on N$_{2}$ as
a function of reduced electric field. (Upper) The red symbols represent
our new experimental measurements, compared with our two-term (black
solid line) and multi-term (blue dashed line) flux drift velocity.
The multi-term flux drift velocity is also compared with the multi-term
bulk (red dashed line) and steady-state Townsend drift velocities
(cyan dotted line). (Lower) Our current experimental measurements
(blue symbols) compared with the Townsend ionisation coefficient calculated
using our multi-term second order approximation using bulk transport
coefficients calculated in the hydrodynamic regime (black solid line),
and the lowest eigenvalue of the time-independent spatially varying
SST solution using a multi-term (red dashed line) and 2-term solution
(blue dotted line). }
\label{fig: N2_WTI_2term}
\end{figure}

In consideration of the differences between our calculations and the
experimental data above, we have investigated the effect of $l_{\textrm{max}}$
on the calculated drift velocity and the Townsend ionisation coefficient,
specifically using a two-term solution. The limitations of the two-term
approximation have been discussed in detail previously (e.g.,~\citep{White2003a})
and our calculations here illustrate some of these differences. Figure~\ref{fig: N2_WTI_2term}
shows our results using two-term and multi-term Boltzmann calculations,
where the agreement with our experimental measurements is improved
by using a two-term solution. The errors decrease from less than 2.8\%
and 14--56\% difference, between our experimental drift velocity
and the Townsend ionisation coefficient for our multi-term calculations,
respectively, to less than 2.6\% and 8--28\% for the two-term calculation
results. While this may initially appear to be counter-intuitive,
in fact it simply reflects that the Biagi~\citep{Biagi_lxcat,Magboltz897}
cross-section database we used was originally engineered for application
with a two-term code to reproduce a selection of the available measured
transport coefficients. For all calculations other than in this figure,
the results presented are from multi-term calculations.

We also note that the neglect of recoil in the inelastic channel in
our solution may have a similar impact on our N$_{2}$ calculations
as for our model cross-section. The mass ratio and rotational threshold
in the model are similar to those for N$_{2}$, and the power transfer
rates show similar behaviour to the model calculations. This observation
may be able to account for some of the underestimation of our calculated
transport coefficients when compared with those from the present experiment.

Transport coefficients are dependent on how the experimental current
trace is analysed~\citep{Robson1991c,WhitePT}, so in figure~\ref{fig: N2_WTI_2term}
we compare the calculated flux, bulk, and steady-state Townsend drift
velocities to the drift velocity extracted from our pulsed Townsend
experiment. It is expected that the differences between the various
possible drift velocities increase with increasing reduced electric
field, with the particle non-conserving ionisation channel increasing
in importance (as shown in the power transfer rates in figures~\ref{fig: N2_EnTrf}
and \ref{fig: N2_EnTrf2}). At the highest measured reduced electric
field of our new experimental data, 360~Td, the difference between
the flux and SST drift velocities is 2.5\%, while a 10\% difference
is calculated between the bulk and flux drift velocities at this field.

For our calculated  Townsend ionisation coefficient we find less than
a 0.1\% difference between the coefficient calculated from a second
order approximation to the Townsend ionisation coefficient using the
bulk time-of-flight coefficients (given in equation~\ref{eq: TOF_Townsend}),
and the direct calculation of the coefficient using our steady-state
Townsend simulation over the range of our experimental values, as
shown in the lower pane of figure~\ref{fig: N2_WTI_2term}.

The neglect of de-excitation processes in the $l\geq1$ terms of the
inelastic collision operator, but inclusion in the $l=0$ term, is
considered in N$_{2}$ to assess their effect on the drift velocity
and the Townsend ionisation coefficient. Unlike with our model calculations
in section~\ref{subsec: Model_approximations}, with only one excitation
channel, in N$_{2}$ the scaling of the ground-state excitations,
when modifying state populations, requires more consideration as the
number of rotational channels significantly populated changes with
the temperature of the neutrals. There are multiple scalings of the
ground-state equation that may be considered (for example, with non-zero
temperature or at 0~K, or with or without degeneracy considered),
so we have assessed the two extreme possibilities combined with setting
the superelastic population to zero for $l\geq1$. We first consider
using the proper ground-state density for $n_{0j}$, calculated using
the neutral temperature, and secondly with no scaling at all (effectively
$n_{0j}=1$): 
\begin{enumerate}
\item When the density of neutrals in the ground state are calculated according
to the temperature of the neutrals, the differences between our standard
293~K calculations of the drift velocity and the Townsend ionisation
coefficient change by less than 4\% and 0.1--31\%, respectively.
The differences between both coefficients decreases with increasing
reduced electric field, as the higher threshold processes with lower
excited-state densities (being neglected) starting to dominate. 
\item For the extreme case of no temperature-dependent scaling or degeneracy
included in the $l\geq1$ equation, when setting $n_{0j}=1$, this
is not equivalent to the case considered in our model calculations
with 0~K in the $l\geq1$ equation of the inelastic operator. In
this case, all of the rotational processes would be weighted equally,
so the resulting transport would be influenced by the number of rotational
cross-sections included in the set. In our current set of more than
40 individual rotational processes, the differences between our standard
calculations and this modified set, for the drift velocity and the
Townsend ionisation coefficient are up to 60\% and 8--100\%, respectively.
The much higher number of neutrals in the ground state for each excitation
channel results in higher momentum exchange with electrons with energies
reduced to $U-U_{th}$, resulting in a reduced mean energy and drift
velocity, and a reduced Townsend ionisation coefficient as sampling
of the ionisation cross-section is delayed to higher reduced electric
fields.
\end{enumerate}
For both of these extreme cases, the neglect of superelastic processes
has an important impact on the calculated transport coefficients,
particularly the Townsend ionisation coefficient, when considering
the $\pm$0.1\% accuracy required in swarm calculations, and even
the 10\% error acceptable in plasma applications~\citep{White2003a}.

\subsubsection{NDC in N$_{2}$\label{subsec: NDC-in-N}}

In molecular nitrogen, an NDC region is present in the measurements
of both Pack and Phelps~\citep{PackPhelps_lx} and Lowke~\citep{Lowke1963}
at 77~K and 77.6~K, respectively. Our calculations and those of
Petrovi{\'c} \emph{et al.}~\citep{Petrovic1984} are in good agreement
with the experimental measurements of these authors. Petrovi{\'c}
\emph{et al.}~\citep{Petrovic1984} also presented drift velocity
calculations at 293~K with and without superelastic collisions, that
our calculations are in similarly good agreement with. An increase
in the drift velocity and the appearance of an NDC region occurs in
the absence of superelastic processes. Highlighted in their work is
the importance of superelastic collisions for both the low-threshold
rotational and vibrational excitation channels. As with our model
cross-section, the temperature dependence of the excitation state
populations in N$_{2}$, and the reduced nett energy transfer due
to inelastic collisions, compared to the 0~K calculations, is responsible
for the absence of NDC at 195~K and 293~K. At 77~K, the decreased
population of the excited states for rotational and vibrational excitations
results in a higher nett energy transfer rate in those channels, and
a corresponding decrease in the drift velocity with increasing reduced
electric field. In figures~\ref{fig: N2_EnTrf} and \ref{fig: N2_EnTrf2}
the power transfer rates for each collision type are given for each
of the temperatures considered. The explicit power transfer due to
the de-excitation processes (in particular rotational excitations,
here grouped into a single line for the figures only) illustrate that
the increased contribution of the superelastic processes at the higher
temperatures is responsible for the decreased range and eventual disappearance
of the NDC region, where the increased mean energy of the swarm changes
the sampled region of the elastic collision frequency, resulting in
an increased drift velocity. The temperature dependence of the NDC
region has been observed previously through the vibrational channel
temperatures in mixtures of 99\% argon and 1\% N$_{2}$~\citep{Dyatko2010},
and our present discussion highlights the same dependence occurring
in a pure gas.

\begin{figure}
\includegraphics[width=0.9\columnwidth]{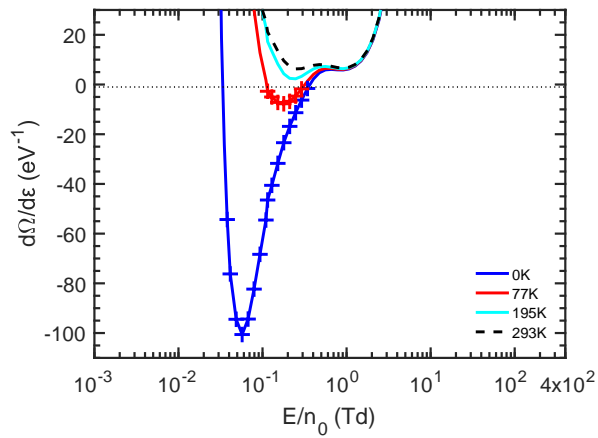}\foreignlanguage{american}{\protect}

\caption{Rate of change of $\Omega$ with mean energy as a function of reduced
electric field for electrons in N$_{2}$. The solid lines show $\partial\Omega/\partial\varepsilon$
calculated from the energy transfer rates for N$_{2}$ at varying
temperatures, where NDC is predicted by
Robson's criterion when $\partial\Omega/\partial\varepsilon<-1$~\citep{Robs84}
(shown by the horizontal dotted line), and
the symbols indicate where NDC is present in our calculated values.}

\label{fig: N2_OmegaField}
\end{figure}

The calculation of $\Omega$~\citep{Robs84} at each temperature
is again a very good predictor for the presence of NDC in molecular
nitrogen. Observed at 0~K and 77~K, figure~\ref{fig: N2_OmegaField}
shows that the prediction of NDC using the criterion $\partial\Omega/\partial\varepsilon<-1$
was consistent with our calculations.

\section{Concluding Remarks\label{sec:Concluding-Remarks}}

In this study we have investigated the temperature dependence of NDC
using a simple model system, alongside that of N$_{2}$. The power
transfer rates in the elastic, inelastic and superelastic channels
show the damping effect of the de-excitation processes on the range
of NDC. With increasing temperatures, the higher proportion of neutral
background gas particles in excited states increases the mean energy
and subsequently suppresses the NDC region that arises from the increasing
elastic momentum-transfer cross-section compared with the (decreasing
importance of the) inelastic channels at those fields. 

To assess the impact of superelastic collisional processes on NDC,
we employed some model (although unphysical) cases, with temperature
dependence during elastic, excitation and de-excitation processes
manipulated, illustrating the importance of the de-excitation process
to the transport coefficients at low reduced electric fields. These
systems also isolated the physical processes responsible for NDC,
with the energy gained by the electron swarm from the de-excitation
channel reducing the range of or eliminating NDC altogether. 

We have also presented calculations of Robson's~\citep{Robs84} criterion
for the presence of NDC using the rate of change of the ratio of the
nett energy exchange of inelastic to elastic collisions, derived using
momentum transfer theory. That criterion predicts well the region
of NDC in all of the model cases considered, as well as the temperature-dependent
NDC region present in N$_{2}$, using only a knowledge of the collision
frequencies.

The effect of anisotropic scattering, for very-low threshold inelastic
processes on the transport coefficients, was assessed using a model
cross-section to replicate the forward-peaked nature of rotational
collisions. The inclusion of an inelastic momentum-transfer cross-section
results in a 5--11\% increase in the drift velocity, and between
a 4--9\% change in the mean energy of the electron swarm for the
temperatures considered in this work.

In the Frost-Phelps differential finite difference form of the inelastic
collision operator utilised in this work, the representation of inelastic
collisions is truncated at zeroth order in the mass ratio, neglecting
the recoil of the neutral particle during an inelastic collision.
The effect of this assumption had been assessed previously and found
to have less than a 0.1\% impact on the calculated transport coefficients
for electrons in H$_{2}$~\citep{WhiteMorrMas2002}. For the model
cross-section considered in this work, however, the inelastic threshold
is more than 20 times lower than the lowest rotational threshold in
H$_{2}$, and the impact of the truncation of the mass ratio for inelastics
was found to have a greater influence on the transport coefficients.
At 0~K recoil accounts for a less than 2\% change in the drift velocity
and mean energy, but this difference increased to over 6\% at room
temperature. To derive the next terms in the mass ratio expansion
for the Frost-Phelps inelastic operator was beyond the scope of the
present work, however should be considered when adjusting cross-sections
derived from swarm transport measurements for processes with very
low thresholds (for example, the derived vibrational cross-sections
for H$_{2}$~\citep{WhiteMorrMas2002,WhiteRobMorrLiNess2007}). 

Finally, we have reported experimental measurements of the drift velocity
and the Townsend ionisation coefficient for electron swarms in gaseous
N$_{2}$ using a pulsed Townsend apparatus. Comparison of our measurements
with some of the other available experimental measurements shows reasonable
agreement, generally within $\pm$10\% for the drift velocity, and
$\pm$5--50\% for the Townsend ionisation coefficient. We have also
presented our calculations using a multi-term Boltzmann solution,
that well reproduce experimental drift velocities and the Townsend
ionisation coefficients at 293~K, but tend to somewhat overestimate
our current experimental measurements. At lower temperatures, we also
tend to underestimate the experimental drift velocities at the lower
reduced electric fields, however addressing this by modification of
the N$_{2}$ cross-sections we employed through a swarm analysis was
outside the scope of this work. Rather, our calculations were used
to illustrate the physical processes associated with NDC and the effect
of temperature on its appearance or absence, with the same dependence
on superelastic populations found as in our model calculations.
\begin{acknowledgments}
The authors would like to thank the Australian Research Council through
its Discovery Program (DP180101655) for financial support. The experimental
study was supported by UNAM-PAPIIT IN 108417 and Conacyt project 240073.
The technical assistance of A. Bustos, G. Bustos and H. Figueroa is
greatly acknowledged. SD would like to acknowledge MPNTR projects
ON171037 and III41011 for support.
\end{acknowledgments}

\appendix

\section{Numerical Considerations\label{sec: Appendix Numerical}}

The theory and solution techniques employed in this study have been
systematically benchmarked against independent Monte Carlo and kinetic
theory solutions. Using model systems, each of the collisional processes
have been validated by comparing against existing solutions for the
hard sphere and Maxwell's models, Reid's ramp and anisotropy models~\citep{Reid1979},
and attachment and ionisation models~\citep{NessRob1986}. When we
investigate these model systems outside of the reduced electric field
and temperature ranges generally considered, we sometimes find problematic
behaviour in their solutions and resulting transport coefficients.
For some real gases we are also able to provoke similar behaviour,
where the electron energy distribution function contains a contribution
from a solution that is not part of the physical solution we expect
to extract, as illustrated below. Generally these problematic regions
are outside standard swarm experimental configurations, but warrant
further investigation nonetheless. 

To assess the effect of the non-physical contribution on our calculations,
for all of our results here we compare with an independent Monte Carlo
solution, discussed in section~\ref{subsec: Monte-Carlo-Technique},
and find that for the conditions required for calculations involving
real gases, and the model system employed here, we reproduce the Monte
Carlo results, with the exception of the NDC region of our model system
where recoil of the neutral during inelastic collisions becomes important,
as discussed in section~\ref{subsec:Benchmark-model}. As part of
our investigation into the origin of these problematic solutions,
we have trialed different boundary conditions from those of Winkler
and collaborators~\citep{Loffhagen1996a} that we usually employ.
The generalised eigenvalue method utilised for our time-of-flight
and steady-state Townsend solutions can be solved using inbuilt Matlab
functions, or a benchmarked inverse power method, where we enforce
strict convergence criteria. Using a high minimum number of iterations
in the inverse power method solver, improvements in the distribution
function are obtained, but the contributions from the non-physical
solution remain.

We illustrate using N$_{2}$ that a discrepancy exists in our simulations
below the experimental $E/n_{0}$ values. For time-of-flight calculations,
when solved as a generalised eigenvalue problem, the lowest temporal
eigenvalue is equivalent to the nett rate coefficient, but is not
explicitly used in calculations of the transport coefficients, given
in section~\ref{subsec: TransCo}. For the steady-state Townsend
simulations, however, the lowest spatial eigenvalue represents the
 Townsend ionisation coefficient extracted from the experimental measurements.
For any configuration with conservative collisions only, both spatial
and temporal eigenvalues should be zero, and we use this, along with
the equivalence of the nett rate coefficient and temporal eigenvalue,
as a consistency check for our solution. 

\begin{figure}
\includegraphics[width=0.9\columnwidth]{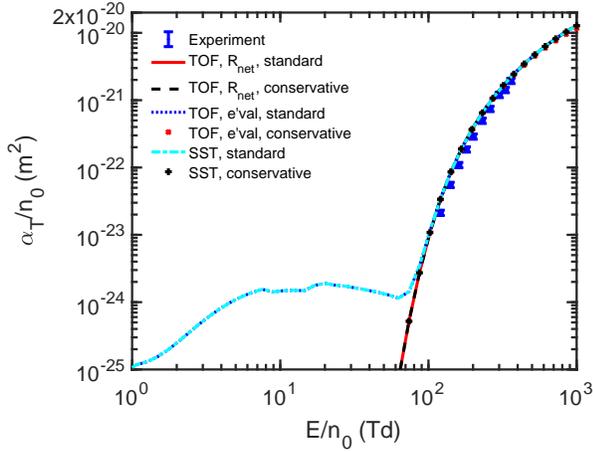}\foreignlanguage{american}{\protect}

\caption{The Townsend ionisation coefficient for electron impact on N$_{2}$
as a function of reduced electric field. The Townsend ionisation coefficient
calculated from a time-of-flight simulation, using a second order
approximation with bulk transport coefficients calculated in the hydrodynamic
regime (solid red line), is compared with (the absolute value of)
the Townsend ionisation coefficient calculated using the hydrodynamic
coefficients with the temporal eigenvalue in place of $R_{\textrm{net}}$
using the standard collision operators (dashed blue line, directly
underneath the cyan line), and the coefficient calculated directly
from our steady-state Townsend simulation (cyan dash-dotted line).
Compared with the standard collision operator representation are our
calculations employing the conservative representation of the collision
operators, using time-of-flight coefficients using $R_{\textrm{net}}$
(black dashed line) and employing the eigenvalue in place of $R_{\textrm{net}}$
(red symbols), and the direct calculation of the Townsend ionisation
coefficient from the steady-state Townsend solution (black symbols).}
\label{fig: N2_TI_numerical}
\end{figure}

Depicted in figure~\ref{fig: N2_TI_numerical} for N$_{2}$ is the
Townsend ionisation coefficient calculated using these equivalent
methods. In the low field region (below our present experimental measurements)
the ionisation cross-section is sampled by the tail region of the
distribution function only, so we would expect to see the Townsend
ionisation coefficient decreasing with decreasing reduced electric
field. Using the bulk transport coefficients from a time-of-flight
simulation in a second order approximation to the spatial rate coefficient
(given in equation~\ref{eq: TOF_Townsend}), our calculations well
reproduce the experimental results, labelled `TOF $R_{\textrm{net}}$,
standard' in figure~ \ref{fig: N2_TI_numerical}.

With the temporal eigenvalue used in place of the equivalent nett
rate coefficient ($R_{\textrm{net}}$ in equation~\ref{eq: TOF_Townsend}),
labelled `TOF e'val, standard' in figure~\ref{fig: N2_TI_numerical},
in the low field region our calculations produce non-zero values for
the Townsend ionisation coefficient, showing an inconsistency in our
time-of-flight simulation. 

In our steady-state Townsend simulation (labelled `SST, standard'
in figure~\ref{fig: N2_TI_numerical}), where the lowest eigenvalue
corresponds directly to the spatial ionisation rate coefficient, we
observe the exact same non-zero contributions at reduced electric
fields below where ionising collisions should be contributing. The
consistency of this non-zero contribution between the two different
simulations suggests a leak of number density stemming from the same
numerical source.

To address these issues, in the following
subsections we consider the numerical representation of the collision
operators, where using an alternative description of the collision
operators results in electron energy distribution functions where
the noise in the tail region is suppressed to satisfactory higher
energies/lower magnitudes. In regions where the distribution function
does not drop a satisfactory amount, all of our results are calculated
using these alternative collision operators and are compared with
an equivalent Monte Carlo solution, see section~\ref{subsec: Monte-Carlo-Technique}.
Under these circumstances we find very good agreement between the
distributions and resulting transport coefficients. This is illustrated
in N$_{2}$ by the Townsend ionisation coefficient calculations labelled
`Conservative' in figure~\ref{fig: N2_TI_numerical}, where using
the time-of-flight coefficients with either $R_{\textrm{net}}$ or
the equivalent temporal eigenvalue, and our steady-state Townsend
code reproduce the expected values. We note that each of the modifications
proposed below reproduce the standard model benchmark systems.

\subsection{Representation of the derivative in the elastic collision operator}

\begin{figure}
\includegraphics[width=0.9\columnwidth]{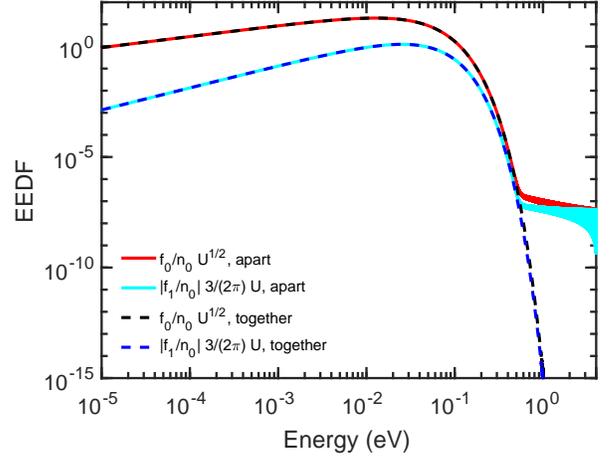}\foreignlanguage{american}{\protect}

\caption{The first two terms of the electron energy distribution function (EEDF):
$U^{1/2}f_{0}/n_{0}$ (eV$^{-1}$) and $|3/(2\pi)Uf_{1}/n_{0}|$ (eV$^{-1/2}$)
as a function of energy for Maxwell's model elastic cross-section
at 293~K under a reduced electric field $E/n_{0}=10^{-4}\,$Td. The
solid lines represent the distributions calculated with the derivative
term in the elastic collision operator separated (labelled `apart'),
while the dashed lines represent the solution with the derivative
term together (labelled `together'). Here the noise in the tail region
of the distribution is suppressed to higher energies so that the distribution
captures a sufficient amount of particles over a sufficient energy
range.}
\label{fig: LowFieldProb_ElasticOp}
\end{figure}

For elastic collisions with non-stationary neutrals, the collision
operator has a single and double derivative term in the $l=0$ expression.
Using the collision operator when the first derivative term is expanded
with the product rule meets all of the existing benchmarks. Similarly,
if we now numerically represent the collision operator as $J_{0}^{\textrm{elas}}\left(f_{0}\right)=-\frac{2m_{e}}{m_{0}}U^{-1/2}\frac{\partial}{\partial U}\left[U^{3/2}\nu_{m}^{\textrm{elas}}(U)\left(f_{0}+\frac{k_{B}T_{0}}{q}\frac{\partial}{\partial U}f_{0}\right)\right]$,
again all benchmarks are met, and for some situations the contribution
of the noise in the tail region of the electron energy distribution
is delayed to higher energies, allowing the solution to capture a
sufficient energy range of the electron swarm. This is illustrated
by an example using Maxwell's elastic model cross-section at 293~K
with $E/n_{0}=10^{-4}$~Td, with the resulting $f_{0}$ and $f_{1}$
terms of the distribution function shown in figure~\ref{fig: LowFieldProb_ElasticOp}. 

This representation is equivalent to a finite volume approach, where
a flux of electron density is moved between energy grid elements,
with conservation of this density enforced. For our illustration in
N$_{2}$ in figure~\ref{fig: N2_TI_numerical}, using the time-of-flight
coefficients and temporal eigenvalue in place of $R_{\textrm{net}}$,
this conservation of density reduces the Townsend ionisation coefficient
calculated at reduced electric fields below where the ionisation channel
has a significant impact on transport, however a non-zero contribution
remains, to be addressed in the following section that considers the
representation of inelastic collisions.

Similar considerations for the representation of the field term may
be taken, representing the derivatives together as other authors employ
(e.g.,~\citep{Loffhagen2002}), or expanding using the product rule
(as in reference~\citep{Trunec2006}), but we have found very little
difference in the resulting distributions or transport coefficients.

\subsection{Conservation of number density in inelastic collisions}

To address the inconsistency between the eigenvalue and nett rate
coefficient, highlighted for N$_{2}$ in figure~\ref{fig: N2_TI_numerical}
above, we reformulate the inelastic collision operators to force conservation
of electron density in a finite-volume style. To calculate the change
in the distribution function $f(U,t)$, due to an inelastic collisional
process $j$, an expression is needed for the change of the number
density over time $\frac{dn}{dt}$, in terms of known quantities,
$\nu^{j}(U)$, and calculable quantities, $f(U)$ or $n(U)$. 

For the flux of particles in and out of each energy bin element to
be conserved, to move (or add for ionising collisions) particles between
pre- and post-collision positions in the energy grid, a movement matrix
$\mathbf{M}$ is applied to move the density of particles in each
volume element surrounding the solution energy grid $U=0,...,U_{\infty}$.
The general form for the change of the distribution function $f_{0}$
due to an inelastic collision $j$ is then given by $J_{0}^{j}(f_{0}(U))=n_{0}\sigma^{*}\left(\frac{2q}{m_{e}}\right)^{1/2}\frac{1}{\sqrt{\mathbf{U}}}\frac{1}{\Delta\mathbf{U}}\mathbf{M}$,
where $n_{0}$ is the neutral number density, $\sigma^{*}=10^{-20}$
scales the cross-section in m$^{2}$ to $\lyxmathsym{\AA}^{2}$, and
$\Delta\mathbf{U}$ are the bin widths (i.e. $\Delta U_{i}=U_{i+1}-U_{i}$).
The energy dependent quantities to the left scale the conserved number
density (contained in $\mathbf{M}$) per energy element per unit time,
to the distribution function modified by the collision $j$. Here,
the conserved quantity is the number density scaled by the collision
frequency for the process $j$, taken to be $\rho(U)=\nu^{j}(U)\sqrt{U}f_{0}(U)$. 

The general form of the movement matrix $\mathbf{M}$ calculates the
density of particles from the pre-scattered energy element $U_{i+1/2}$,
with associated bin $U_{i+1}-U_{i}$, that move into each of the energy
bins in the post-collision scattering region $\left(U_{i+1}\pm U_{th}\right)-\left(U_{i}\pm U_{th}\right)$,
and varies for inelastic, superelastic and ionising collisions. For
an expression for the flux of the density of particles in and out
of each bin, we consider the general case where the boundaries of
the pre- or post-scattered energy bin lie within (not at the edges
of) a volume element surrounding a point in the solution grid (i.e.
for an element of the solution grid to be scattered, $U_{i+1}-U_{i},$
and post-scattered energy bin with left and right boundaries $U_{r}-U_{l}$,
taken to lie within a grid element $U_{j+1}-U_{j}$, $U_{l}=U_{i}-U_{th}>U_{j}$
and $U_{r}=U_{i+1}-U_{th}<U_{j+1}$). The number density between a
left $l$ and right $r$ boundary can be calculated using the density
at the bin edges, and here we choose to linearly interpolate the distribution
function. For a bin $U_{r}-U_{l}$, the density of particles, scaled
by the collision frequency for the process $j$, taken to be $\rho(U)=\nu^{j}(U)\sqrt{U}f_{0}(U)$,
is given by $\int_{U_{l}}^{U_{r}}\rho(U)dU=\int_{U_{l}}^{U_{r}}\nu^{j}(U)\sqrt{U}f_{0}(U)dU$.
When $\rho(U)$ is assumed to be linearly spread across each bin width,
the expression for $\rho(U)$ at some energy $U$ is given by: $\rho(U)=\frac{\rho_{i+1}-\rho_{i}}{\Delta U_{i}}U+\frac{U_{i+1}\rho_{i}-U_{i}\rho_{i+1}}{\Delta U_{i}}$,
so that 
\begin{eqnarray*}
\int_{U_{l}}^{U_{r}}\rho(U)dU & = & \Bigg[\left(\frac{1}{2}U^{2}-UU_{i}\right)\frac{\rho_{i+1}}{\Delta U_{i}}\\
 &  & -\left(\frac{1}{2}U^{2}-UU_{i+1}\right)\frac{\rho_{i}}{\Delta U_{i}}\Bigg]_{U_{l}}^{U_{r}}\\
 & = & \left[\frac{1}{2}\left(U_{r}+U_{l}\right)-U_{i}\right]\frac{\Delta U_{rl}}{\Delta U_{i}}\rho_{i+1}\\
 &  & -\left[\frac{1}{2}\left(U_{r}+U_{l}\right)-U_{i+1}\right]\frac{\Delta U_{rl}}{\Delta U_{i}}\rho_{i},
\end{eqnarray*}
 where $\Delta U_{rl}=U_{r}-U_{l}$. When the left and right boundary
fall on elements of the solution grid, this expression reduces to
$\int_{U_{l}}^{U_{r}}\rho(U)dU=\frac{1}{2}\left(U_{i+1}-U_{i}\right)\left(\rho_{i+1}+\rho_{i}\right)$.
For any valid solution, a fine enough grid must be taken over the
energy range considered. Namely, the bin widths are sufficiently small
so that the choice of linear interpolation does not affect the solution
and a convergence criterion in the number of grid elements is met.
For particles scattered (post-collision) outside the range of the
simulation grid, to maintain conservation of number density for a
physically realistic system with $U\geq0$, those particles must remain
in the system and are allocated to the lowest (or highest for superelastic
collisions) bin of the simulation. For superelastic collisions these
expressions remain the same, with particles lost from the $U_{i+1}-U_{i}$
bin and gained in the $U_{r}-U_{l}=\left(U_{i+1}+U_{th}\right)-\left(U_{i}+U_{th}\right)$
bin, and $\rho$ is evaluated for the appropriate collision frequency.

In the case of the benchmarks involving conservative collisions only,
the new representation of the inelastic collision operator, when solved
as a generalised eigenvalue problem~\citep{Boyle2015thesis}, gives
a numerically zero eigenvalue, consistent with the equivalent zero
nett rate coefficient.

For ionising collisions, the collision operator takes a form similar
to the inelastic operator for ground to excited state collisions,
except that the two post-collision electrons must be allocated to
a shifted energy bin ($U_{r}-U_{l}$) based on the particular energy
sharing fraction describing the process. The existing ionisation benchmarks
of Ness and Robson~\citep{NessRob1986} are also satisfied with this
conservative representation of the collision operators. 

We also note that since it is the $l=0$ equations that correspond
to the change in number density, similar representations of the $l\geq1$
collision operators have little effect on the eigenvalue of the solution
and corresponding transport coefficients, with generally less than
a $10^{-4}$\% difference.

Using this alternate representation for the excitation and ionisation
collisions all of the collisional benchmarks we consider are satisfied,
and importantly the nett rate coefficient and eigenvalue from the
time-of-flight simulations are consistent. 

For N$_{2}$, using these finite-volume style conservative collision
operators, we find the  Townsend ionisation coefficient, below where
ionisation contributes, drops off as we would expect with decreasing
reduced electric fields. The Townsend ionisation coefficient calculated
using the second-order hydrodynamic coefficients with the eigenvalue
in place of $R_{\textrm{net}}$ is shown in figure~\ref{fig: N2_TI_numerical}
(labelled `TOF e'val, conservative') to be consistent with the coefficient
calculated using the nett rate coefficient (labelled \foreignlanguage{american}{`}TOF
$R_{\textrm{net}}$, conservative'). Similarly, the Townsend ionisation
coefficient calculated directly from a steady-state Townsend simulation
using the conservative operators (labelled `SST, conservative') shows
only the contribution from ionising collisions, and is consistent
with the experimental measurements, as shown in figure~\ref{fig: N2_WTI}.

\section{Measured transport coefficients in N$_{2}$\label{sec: ExpTransCo}}

\selectlanguage{american}%
\begin{table}[H]
\caption{Variation of the present measured drift velocity and the Townsend
ionisation coefficient with reduced electric field $E/n_{0}$ for
electron impact on N$_{2}$.}
\label{Table: Jaime-Measurements}

{\footnotesize{}}%
\begin{longtable}[c]{|c|c|c|}
\hline 
{\footnotesize{}$E/n_{0}$ (Td)} & \multicolumn{1}{c|}{{\footnotesize{}$W$ (m/s) $\pm$2.2\%}} & \multicolumn{1}{c|}{{\footnotesize{}$\alpha_{T}/n_{0}\ (\textrm{m}^{2})$ $\pm$9\%}}\tabularnewline
\hline 
\hline 
0.65\selectlanguage{american}%
 & 
3750\selectlanguage{american}%
 & 
\tabularnewline
\hline 
0.7\selectlanguage{american}%
 & 
3840\selectlanguage{american}%
 & 
\selectlanguage{american}%
\tabularnewline
\hline 
0.8\selectlanguage{american}%
 & 
4020\selectlanguage{american}%
 & 
\selectlanguage{american}%
\tabularnewline
\hline 
0.9\selectlanguage{american}%
 & 
4170\selectlanguage{american}%
 & 
\selectlanguage{american}%
\tabularnewline
\hline 
1\selectlanguage{american}%
 & 
4350\selectlanguage{american}%
 & 
\tabularnewline
\hline 
1.2\selectlanguage{american}%
 & 
4620\selectlanguage{american}%
 & 
\selectlanguage{american}%
\tabularnewline
\hline 
1.4\selectlanguage{american}%
 & 
4870\selectlanguage{american}%
 & 
\selectlanguage{american}%
\tabularnewline
\hline 
1.6\selectlanguage{american}%
 & 
5160\selectlanguage{american}%
 & 
\selectlanguage{american}%
\tabularnewline
\hline 
1.8\selectlanguage{american}%
 & 
5460\selectlanguage{american}%
 & 
\selectlanguage{american}%
\tabularnewline
\hline 
2\selectlanguage{american}%
 & 
5750\selectlanguage{american}%
 & 
\selectlanguage{american}%
\tabularnewline
\hline 
2.3\selectlanguage{american}%
 & 
6210\selectlanguage{american}%
 &
\selectlanguage{american}%
\tabularnewline
\hline 
2.6\selectlanguage{american}%
 & 
6670\selectlanguage{american}%
 & 
\selectlanguage{american}%
\tabularnewline
\hline 
3\selectlanguage{american}%
 & 
7310\selectlanguage{american}%
 & 
\selectlanguage{american}%
\tabularnewline
\hline 
3.3\selectlanguage{american}%
 &
7760\selectlanguage{american}%
 &
\selectlanguage{american}%
\tabularnewline
\hline 
3.6\selectlanguage{american}%
 & 
8230\selectlanguage{american}%
 & 
\selectlanguage{american}%
\tabularnewline
\hline 
4\selectlanguage{american}%
 & 
8850\selectlanguage{american}%
 & 
\selectlanguage{american}%
\tabularnewline
\hline 
4.5\selectlanguage{american}%
 & 
9650\selectlanguage{american}%
 & 
\selectlanguage{american}%
\tabularnewline
\hline 
5\selectlanguage{american}%
 & 
10400\selectlanguage{american}%
 & 
\selectlanguage{american}%
\tabularnewline
\hline 
5.5\selectlanguage{american}%
 & 
11100\selectlanguage{american}%
 & 
\selectlanguage{american}%
\tabularnewline
\hline 
6\selectlanguage{american}%
 & 
11900\selectlanguage{american}%
 & 
\selectlanguage{american}%
\tabularnewline
\hline 
6.5\selectlanguage{american}%
 & 
12600\selectlanguage{american}%
 & 
\selectlanguage{american}%
\tabularnewline
\hline 
7\selectlanguage{american}%
 & 
13300\selectlanguage{american}%
 & 
\selectlanguage{american}%
\tabularnewline
\hline 
8\selectlanguage{american}%
 & 
14600\selectlanguage{american}%
 & 
\selectlanguage{american}%
\tabularnewline
\hline 
9\selectlanguage{american}%
 & 
16000\selectlanguage{american}%
 & 
\selectlanguage{american}%
\tabularnewline
\hline 
10\selectlanguage{american}%
 & 
17200\selectlanguage{american}%
 & 
\selectlanguage{american}%
\tabularnewline
\hline 
12\selectlanguage{american}%
 & 
19800\selectlanguage{american}%
 & 
\selectlanguage{american}%
\tabularnewline
\hline 
14\selectlanguage{american}%
 & 
22100\selectlanguage{american}%
 & 
\selectlanguage{american}%
\tabularnewline
\hline 
16\selectlanguage{american}%
 & 
24400\selectlanguage{american}%
 & 
\selectlanguage{american}%
\tabularnewline
\hline 
18\selectlanguage{american}%
 & 
26700\selectlanguage{american}%
 & 
\selectlanguage{american}%
\tabularnewline
\hline 
20\selectlanguage{american}%
 &
28900\selectlanguage{american}%
 &
\selectlanguage{american}%
\tabularnewline
\hline 
23\selectlanguage{american}%
 & 
32100\selectlanguage{american}%
 &
\selectlanguage{american}%
\tabularnewline
\hline 
26\selectlanguage{american}%
 & 
35200\selectlanguage{american}%
 & 
\selectlanguage{american}%
\tabularnewline
\hline 
30\selectlanguage{american}%
 & 
39200\selectlanguage{american}%
 & 
\selectlanguage{american}%
\tabularnewline
\hline 
33\selectlanguage{american}%
 & 
42100\selectlanguage{american}%
 & 
\selectlanguage{american}%
\tabularnewline
\hline 
36\selectlanguage{american}%
 & 
45000\selectlanguage{american}%
 &
\selectlanguage{american}%
\tabularnewline
\hline 
40\selectlanguage{american}%
 & 
48600\selectlanguage{american}%
 & 
\selectlanguage{american}%
\tabularnewline
\hline 
45\selectlanguage{american}%
 & 
52900\selectlanguage{american}%
 & 
\selectlanguage{american}%
\tabularnewline
\hline 
50\selectlanguage{american}%
 & 
57200\selectlanguage{american}%
 & 
\selectlanguage{american}%
\tabularnewline
\hline 
55\selectlanguage{american}%
 & 
61500\selectlanguage{american}%
 & 
\selectlanguage{american}%
\tabularnewline
\hline 
60\selectlanguage{american}%
 & 
65800\selectlanguage{american}%
 & 
\selectlanguage{american}%
\tabularnewline
\hline 
65\selectlanguage{american}%
 & 
69900\selectlanguage{american}%
 & 
\selectlanguage{american}%
\tabularnewline
\hline 
70\selectlanguage{american}%
 & 
73900\selectlanguage{american}%
 & 
\selectlanguage{american}%
\tabularnewline
\hline 
80\selectlanguage{american}%
 & 
82300\selectlanguage{american}%
 & 
\selectlanguage{american}%
\tabularnewline
\hline 
90\selectlanguage{american}%
 & 
91900\selectlanguage{american}%
 & 
\selectlanguage{american}%
\tabularnewline
\hline 
100\selectlanguage{american}%
 & 
100000\selectlanguage{american}%
 & 
\selectlanguage{american}%
\tabularnewline
\hline 
120\selectlanguage{american}%
 & 
119000\selectlanguage{american}%
 & 
2.13\foreignlanguage{american}{{\footnotesize{}$\times10^{-23}$}}\selectlanguage{american}%
\tabularnewline
\hline 
140\selectlanguage{american}%
 & 
137000\selectlanguage{american}%
 & 
5.53\foreignlanguage{american}{{\footnotesize{}$\times10^{-23}$}}\selectlanguage{american}%
\tabularnewline
\hline 
160\selectlanguage{american}%
 & 
157000\selectlanguage{american}%
 & 
1.09\foreignlanguage{american}{{\footnotesize{}$\times10^{-22}$}}\selectlanguage{american}%
\tabularnewline
\hline 
180\selectlanguage{american}%
 & 
176000\selectlanguage{american}%
 & 
1.86\foreignlanguage{american}{{\footnotesize{}$\times10^{-22}$}}\selectlanguage{american}%
\tabularnewline
\hline 
200\selectlanguage{american}%
 & 
195000\selectlanguage{american}%
 & 
2.89\foreignlanguage{american}{{\footnotesize{}$\times10^{-22}$}}\selectlanguage{american}%
\tabularnewline
\hline 
230\selectlanguage{american}%
 & 
220000\selectlanguage{american}%
 & 
4.96\foreignlanguage{american}{{\footnotesize{}$\times10^{-22}$}}\selectlanguage{american}%
\tabularnewline
\hline 
260\selectlanguage{american}%
 & 
246000\selectlanguage{american}%
 & 
7.44\foreignlanguage{american}{{\footnotesize{}$\times10^{-22}$}}\selectlanguage{american}%
\tabularnewline
\hline 
300\selectlanguage{american}%
 & 
274000\selectlanguage{american}%
 & 
1.2\foreignlanguage{american}{{\footnotesize{}$\times10^{-21}$}}\selectlanguage{american}%
\tabularnewline
\hline 
330\selectlanguage{american}%
 & 
295000\selectlanguage{american}%
 & 
1.42\foreignlanguage{american}{{\footnotesize{}$\times10^{-21}$}}\selectlanguage{american}%
\tabularnewline
\hline 
360\selectlanguage{american}%
 & 
323000\selectlanguage{american}%
 & 
1.94\foreignlanguage{american}{{\footnotesize{}$\times10^{-21}$}}\selectlanguage{american}%
\tabularnewline
\hline 
\end{longtable}{\footnotesize\par}
\end{table}
\bibliographystyle{apsrev4-1}
\bibliography{N2}

\end{document}